\documentclass[12pt]{article}%
\usepackage{amsfonts}
\usepackage{physics}
\usepackage{bm}
\usepackage{bbm}
\usepackage{subfiles}
\usepackage{amsmath}
\usepackage{graphicx}
\usepackage{hyperref}
\usepackage{verbatim}
\usepackage{color}
\usepackage{tikz}
\usepackage{pgfplots}
\usepgfplotslibrary{fillbetween}
\usetikzlibrary{calc,decorations.markings,arrows,decorations.pathmorphing}
\usepackage{enumerate}
\usepackage{amssymb}
\usepackage{authblk}
\usepackage[hang,flushmargin]{footmisc}
\usepackage{lipsum}
\usepackage[font=footnotesize]{caption}
\usepackage{subcaption}
\usepackage{empheq}
\allowdisplaybreaks[3]

\csname @addtoreset\endcsname{equation}{section}
 \newcommand{\badat}{\begin{alignedat}}
 \newcommand{\eadat}{\end{alignedat}}

\usepackage{csquotes}
\usepackage[backend=biber, hyperref,backref,backrefstyle=none,style=numeric-comp, sorting=none,maxbibnames=5,url=false]{biblatex}

\DefineBibliographyStrings{english}{%
	backrefpage = {cited on page},%
	backrefpages = {cited on pages}%
}
\DeclareFieldFormat{doi}{%
	\hfil\penalty90\hfilneg\space DOI\addcolon\addnbspace
	\ifhyperref
	{\href{https://doi.org/#1}{\nolinkurl{#1}}}
	{\nolinkurl{#1}}}

\def\be{\begin{eqnarray}}
\def\ee{\end{eqnarray}}
\def\beann{\begin{eqnarray*}}
\def\eeann{\end{eqnarray*}}
\def\beq{\begin{equation}}
\def\eeq{\end{equation}}
\def\ba{\begin{array}}
\def\ea{\end{array}}
\def\ben{\begin{enumerate}}
\def\een{\end{enumerate}}
\def\bea{\begin{eqnarray}}
\def\eea{\end{eqnarray}}

\def\5{\bar }
\def\6{\partial }
\def\7{\hat }
\def\4{\tilde }

\def\cD{\mathcal{D}}
\def\cE{\mathcal{E}}

\def\cH{\mathcal{H}}

\def\cL{\mathcal{L}}

\newcommand{\LL}{\mathcal{L}}

\newcommand{\LLpm}{{\mathcal{L}^\pm}}
\newcommand{\HHpm}{H^\pm}

\renewcommand{\d}{\partial}

\renewcommand{\tilde}{\widetilde}
\renewcommand{\hat}{\widehat}

\begin{document}

\title{\vspace{-70pt} \Large{\sc Nonaxisymmetric (2+1) black holes with Dym boundary conditions}\vspace{10pt}}
\author[a,b]{\normalsize{Miguel Pino}\footnote{\href{mailto:miguel.pino.r@usach.cl}{miguel.pino.r@usach.cl}}}
\author[a]{\normalsize{Francisco Reyes}\footnote{\href{mailto:francisco.reyes.z@usach.cl}{francisco.reyes.z@usach.cl}}}

\affil[a]{\footnotesize\textit{Departamento de F\'isica, Universidad de Santiago de Chile, Avenida V\'ictor Jara 3493, Santiago, Chile.}}
\affil[b]{\footnotesize\textit{Center for Interdisciplinary Research in Astrophysics and Space Exploration (CIRAS),
Universidad de Santiago de Chile, Avenida Libertador Bernardo O’Higgins 3363, Santiago, Chile.}}

\date{}

\maketitle
\thispagestyle{empty}
\begin{abstract}
  We present the spectrum of stationary black hole solutions associated with the first nontrivial member of the Dym family of boundary conditions. The analysis combines the holonomy conditions provided by the Chern-Simons formulation of AdS$_3$ gravity with the periodic solutions of the stationary Dym equation. The resulting spacetimes are characterized by two positive real parameters, two positive integers, and a relative phase, generically describing stationary, nonaxisymmetric black holes.
\end{abstract}

\newpage

\begin{small}
{\addtolength{\parskip}{-2pt}
 \tableofcontents}
\end{small}
\thispagestyle{empty}
\newpage

\section{Introduction}

Motivated by the study of gravitational dualities beyond the standard AdS/CFT correspondence, a variety of boundary conditions for the gravitational field have been explored in the literature~\cite{Barnich:2010eb,Guica:2008mu,Son:2008ye,Hartnoll:2018xxg,Hawking:2016sgy,Hawking:2016msc,Donnay:2015abr,Donnay:2016ejv,Giribet:2021zie,Giribet:2023xom,Kehagias:2016zry,Bonga:2020fhx,Enriquez-Rojo:2020miw,Enriquez-Rojo:2021blc,Pasterski:2023ikd,Donnay:2023mrd,Prema:2021sjp,Troessaert:2017jcm}. Three-dimensional gravity with a negative cosmological constant has proven to be a particularly fruitful arena for such exploration, owing to its relative simplicity. Several proposals~\cite{Henneaux:2002wm,Liu:2009kc,Henneaux:2011hv,Henneaux:2010fy,Compere:2013bya,Troessaert:2013fma,Avery:2013dja,Afshar:2015wjm,Grumiller:2016pqb,Valcarcel:2018kwd,Henneaux:2019sjx,Alessio:2020ioh,Ojeda:2019xih,Afshar:2016wfy,Perez:2016vqo,Fuentealba:2020zkf,Cardenas:2025qqi,Lara:2024cie,Cardenas:2021vwo,Ozer:2021wtx,Ozer:2019nkv} have extended the well-known Brown-Henneaux boundary conditions~\cite{Brown:1986nw}, for which the asymptotic symmetries span two copies of the Virasoro algebra with central charge $c=3\ell/2G$, where $\ell$ represents the AdS radius and $G$ is Newton's constant. One approach to modifying these boundary conditions is to exploit the freedom in choosing the boundary behavior of the lapse and shift functions \cite{Banados:2004nr,Bunster:2014mua}. This path is most naturally expressed in the Chern-Simons formulation~\cite{Achucarro:1986uwr,Witten:1988hc}, where the dynamical fields are two $\mathfrak{sl}(2,\mathbb{R})$ connections $\mathcal A^\pm$. The radial coordinate $\rho$ can be removed by a suitable gauge choice,  \begin{align}\label{A}
    \mathcal A^\pm = b_\pm^{-1} a^\pm b_\pm + b_\pm^{-1} d b_\pm \,,
\end{align}
where $b_\pm$ are taken to capture the entire radial dependence of the fields. The boundary conditions are then given by~\cite{Coussaert:1995zp}
\begin{subequations}\label{a}
  \begin{align}
    a_\phi^\pm &= L_{\pm 1} - \frac{12\pi}{c}\,\mathcal{L}^\pm L_{\mp 1}\,,\label{aphi} \\
    a_t^\pm &= \pm\frac{1}{\ell} \left( \mu^\pm a_\phi^\pm \mp \partial_\phi \mu^\pm L_0 + \frac12 \partial_\phi^2 \mu^\pm L_{\mp 1} \right)\,,\label{at}\\
    a_\rho^\pm &= 0\,,\label{arho} 
  \end{align}
\end{subequations}
where $t$ and $\phi$ stand for the temporal and angular coordinates, respectively, while the generators $\{L_1,L_{-1},L_0\}$ span $\mathfrak{sl}(2,\mathbb{R})$. The functions $\mathcal{L}^\pm(t,\phi)$ represent the dynamical modes, and $\mu^\pm(t,\phi)$ encode the boundary behavior of the lapse and shift functions. In this formulation, the zero-curvature condition, equivalent to Einstein's equations, reduces to  
\begin{equation}\label{eommu}
    \pm \ell\,\partial_t \mathcal{L}^\pm
    = 2 \mathcal{L}^\pm \,\partial_\phi \mu^\pm
      + \mu^\pm \,\partial_\phi \mathcal{L}^\pm
      - \frac{c}{24\pi}\,\partial_\phi^3 \mu^\pm \,.
\end{equation}

The $\mu^\pm$ can be chosen freely, provided that the resulting boundary conditions are consistent. One such possibility is to choose $\mu^\pm$ as suitably constructed functionals of $\mathcal{L}^\pm$, for which the evolution equations \eqref{eommu} reduce to members of integrable hierarchies~\cite{Perez:2016vqo,Ojeda:2019xih,Lara:2024cie}. Implications and extensions of these integrable cases have been analyzed in \cite{Fuentealba:2017omf,Gonzalez:2018jgp,Ojeda:2020bgz,Dymarsky:2020tjh,Cardenas:2021vwo,Cardenas:2024hah,Cardenas:2025qqi,Erices:2019onl,Adami:2025pfk}.

A particular family of functionals for $\mu^\pm$, introduced in~\cite{Lara:2024cie}, yields the Dym hierarchy of integrable equations~\cite{Kruskal1975} as the evolution equations. The construction is based on an expansion of $\mu^\pm$ in negative powers of the Brown-Henneaux central charge,  
\begin{equation}
	\label{expansionmu}
	\mu^\pm=1+\sum_{n=1}^{N}\left( \frac{24 \pi}{c}\right)^{N-n+1} \fdv{\HHpm_n}{\LLpm}\,,
\end{equation}
where $H_n^\pm$ denotes the conserved charges of the Dym hierarchy (see Appendix \ref{cargasDym}), up to an arbitrary positive integer $N$.  

An important requirement for any admissible boundary condition is that it must accommodate physically meaningful solutions, such as black holes. As shown in \cite{Lara:2024cie}, configurations with constant $\cL^\pm$ correspond to black hole spacetimes, extending the well-known BTZ geometry \cite{Banados:1992wn,Banados:1992gq}. The present work aims to characterize the spectrum of stationary black hole solutions compatible with the Dym boundary conditions.

Stationary nonaxisymmetric black holes in three dimensions, often referred to as black flowers, are comparatively scarce. In several known examples, the angular dependence is described by arbitrary periodic functions, or the gravitational field equations reduce, for a suitable ansatz, to linear differential equations \cite{Troessaert:2015syk,Alkac:2016xlr,Afshar:2016wfy}. Configurations whose angular profile is instead determined by a nonlinear stationary integrable equation are considerably less common, with notable examples arising in connection with integrable hierarchies \cite{Ojeda:2019xih,Dymarsky:2020tjh,Cardenas:2025qqi}. The Dym boundary conditions provide a new realization of this latter mechanism: the angular dependence of the geometry is constrained by the nonlinear stationary Dym equation, and the existence of smooth black hole solutions requires finding nontrivial periodic solutions of this equation.

Our analysis is carried out in the Chern-Simons formulation, where black holes can be defined in the Euclidean framework~\cite{PhysRevD.51.622, Banados:1998ta,Bunster:2014mua}. In particular, we impose two conditions:
\begin{enumerate}
    \item The Euclidean manifold must have the topology of a solid torus.
    \item The corresponding solution to the Dym equation must satisfy periodic boundary conditions in the angular coordinate $\phi$.
\end{enumerate}

We demonstrate the existence of solutions to the $N=1$ Dym equation fulfilling these requirements, representing nonaxisymmetric black hole spacetimes.  

The paper is organized as follows. In Sec. \ref{Euclidean BH} we derive the conditions that the functions $\mathcal{L}^\pm$ must satisfy in order for the corresponding Euclidean manifold to describe a solid torus. Section \ref{Periodic Solution} is devoted to the construction of periodic solutions to the Dym equation. In Sec. \ref{Metric} we study the metric properties of the resulting black hole. We conclude with remarks regarding the higher-order equations in the Dym family.

\section{Holonomies}\label{Euclidean BH}

Following~\cite{PhysRevD.51.622, Banados:1998ta,Bunster:2014mua}, we define a Euclidean black hole as a manifold with the topology of a solid torus. An advantage of the Chern-Simons formalism is that it allows to probe the topology of the manifold using holonomies $\mathrm{Hol}_\gamma$, where $\gamma$ is a closed curve in the manifold. Hence, for a black hole configuration, the holonomy along the temporal cycle should be trivial, while nontrivial in the angular cycle. In what follows, these two conditions are translated into requirements for the functions $\cL^\pm$.

First, consider the angular holonomy
\begin{equation}
  \mathrm{Hol}_\phi^\pm
  =
  \mathcal{P}\exp\left[
  \int_0^{2\pi}\mathcal A_\phi^\pm\,d\phi
  \right]\,.
\end{equation}
Since the connection is flat on shell, the trace of the holonomy is invariant under continuous deformations of the curve that preserve its homotopy class. Therefore, a sufficient condition for the angular cycle to have nontrivial holonomy is
\begin{equation}
  \Tr\left[\mathrm{Hol}_\phi^\pm\right]\neq 2\,.
\end{equation}

We can find a simple sufficient condition ensuring the above relation, even without knowing the explicit form of the solutions for $\LL$. Consider only the $+$ sector, since analogous arguments apply to the $-$ sector. We proceed to calculate the holonomy by means of the associated linear problem \cite{Balog:1997zz,Sheikh-Jabbari:2014nya}
\begin{equation}\label{linear}
  \frac{dU(\phi)}{d\phi}
  =
  a_\phi^+ U(\phi),
  \qquad
  U(0)=\mathbb{I}\,.
\end{equation}
The auxiliary connection $a_\phi^+$ can be used instead of $\mathcal A_\phi^+$, since the two are related by a gauge transformation and the trace of the holonomy is gauge invariant. Once the solution to \eqref{linear} is found, the angular holonomy is, by definition,
\begin{equation}
  \mathrm{Hol}_\phi^+=U(2\pi)\,.
\end{equation}

We now analyze the solution for the matrix $U(\phi)$ through the corresponding differential equations. For this purpose, assume that it takes the form
\begin{equation}
  U(\phi)=
  \begin{bmatrix}
    u_1(\phi) & v_1(\phi)\\
    u_2(\phi) & v_2(\phi)
  \end{bmatrix}.
\end{equation}
The linear problem \eqref{linear} implies the identifications $u_2=u_1'$ and $v_2=v_1'$, together with the following differential equations:\footnote{For concreteness, we use the representation
\begin{equation}
  L_{1}=
  \begin{bmatrix}
    0&1\\
    0&0
  \end{bmatrix},
  \qquad
  L_{0}=
  \begin{bmatrix}
    -1/2&0\\
    0&1/2
  \end{bmatrix},
  \qquad
  L_{-1}=
  \begin{bmatrix}
    0&0\\
    -1&0
  \end{bmatrix}.
\end{equation}}
\begin{equation}\label{Holode}
  u_1''=\frac{12\pi}{c}\mathcal{L}\,u_1,
  \qquad
  v_1''=\frac{12\pi}{c}\mathcal{L}\,v_1\,.
\end{equation}
Furthermore, the initial condition $U(0)=\mathbb{I}$ implies
\begin{equation}\label{Holinitial}
  u_1(0)=1,\qquad
  v_1(0)=0,\qquad
  u_1'(0)=0,\qquad
  v_1'(0)=1\,.
\end{equation}

If $\mathcal{L}(\phi)>0$ for all $\phi$, then the differential equations \eqref{Holode}, together with the initial conditions \eqref{Holinitial}, imply that $u_1$, $v_1$, $u_1'$, and $v_1'$ are strictly increasing functions. In particular,
\begin{equation}
  u_1(2\pi)>1,
  \qquad
  v_1'(2\pi)>1\,.
\end{equation}
Therefore,
\begin{equation}
  \Tr\left[\mathrm{Hol}_\phi^+\right]
  =
  u_1(2\pi)+v_1'(2\pi)
  >2\,.
\end{equation}
Hence, the condition $\mathcal{L}(\phi)>0$ is sufficient to ensure that the angular holonomy belongs to the hyperbolic conjugacy class of $SL(2,\mathbb{R})$ associated with black hole solutions~\cite{Martinec:1998wm}.

Now consider the holonomy along the thermal cycle. In the Euclidean formulation, where for stationary configurations $\mathcal A_\tau^\pm=i\mathcal A_t^\pm$, the relevant holonomy is
\begin{equation}
  \mathrm{Hol}_\tau^\pm=\mathcal{P}\exp\left[ \int_0^{\beta^\pm}\mathcal A_\tau^\pm d\tau \right]\,,
\end{equation}
where the period $\beta^\pm$ stands for the inverse temperature. For stationary configurations the holonomy is straightforwardly evaluated. Notably, the choice
\begin{equation}\label{beta}
 \beta^\pm=\frac{2 \pi \ell} {\sqrt{\frac{48\pi}{c}  \cL^\pm {\mu^\pm}^2+ (\d_\phi \mu^\pm)^2-2\mu^\pm \d_\phi^2 \mu^\pm}}\,,
\end{equation}
renders the holonomy trivial $\mathrm{Hol}_\tau^\pm=-\mathbb{I}$. Although it seems to be a function of the angle $\phi$, $\beta^\pm$ is actually a constant since  
\begin{equation}\label{delta beta}
  \d_\phi \beta^\pm=-\frac{48 \pi^2 \ell}{c} \frac{\mu^\pm (2 \cL^\pm \d_\phi \mu^\pm+\d_\phi\cL^\pm\mu^\pm-\frac{c}{24\pi}\d_\phi^3 \mu^\pm)} {(\frac{48\pi}{c}  \cL^\pm {\mu^\pm}^2+ (\d_\phi \mu^\pm)^2-2\mu^\pm \d_\phi^2 \mu^\pm)^{3/2}}\,,
\end{equation}
whose numerator is proportional to the stationary equation of motion \eqref{eommu}, and then vanishes for time-independent on-shell configurations. Hence, for \emph{any} stationary solution of Eq. \eqref{eommu}, there is a constant $\beta^\pm$ that trivializes the temporal holonomy. Since in general it might be a complex number, the triviality of the temporal holonomy translates into the simple condition of a real and positive $\beta^\pm$ given by \eqref{beta}.

In summary, the condition $\LL^\pm(\phi) > 0$ for all values of $\phi$, and the existence of a real and positive $\beta^\pm$, ensure that the topology of the manifold corresponds to a solid torus, thereby characterizing black hole configurations. The periodic solutions constructed in the next section fulfill both conditions.

\section{Periodic Solutions}\label{Periodic Solution}

Another necessary condition to describe a stationary black hole is that the corresponding solution of the Dym equations must obey periodic boundary conditions in the angular coordinate $\phi$. Although explicit solutions to Dym equations are scarce~\cite{Fuchssteiner_1992,Mokhtari2011,SolitonsHD, Novikov1999,RoyChowdhuryMukherjee1984,Bordag:1995dp,alezGaxiola2018ITERATIVEMF,XIAO2019123248,Assabaai2022ExactSO,LI2020106276,Li2017}, analytical insight reinforced by numerical methods is sufficient for the purposes of this work. We outline the construction of periodic and time-independent solutions.

For simplicity, we will focus on the case $N=1$, for which \eqref{eommu} becomes the stationary Dym equation
\begin{equation}\label{stationary_dym}
  \partial_\phi \cL^\pm-\partial_\phi^3 \left(\frac{1}{2\sqrt{\cL^\pm}}    \right)=0\,,
\end{equation}
where the functions $\cL^\pm$ depend solely on the angular coordinate $\cL^\pm(\phi)$. We drop the $\pm$ signs in the following expressions and will reinstate them when necessary. 

\begin{figure}[h]
  \centering
  \begin{center}
      \begin{tikzpicture}[scale=0.75, every node/.style={scale=0.8}]
        \def\x{5} 
        \def\y{1} 
        
        \def\A{2} 
        \def\B{1.8945} 
        
    \def\um{(-sqrt(\B^2-\A)+\B)/\A} 
    \def\up{(sqrt(\B^2-\A)+\B)/\A} 
    
    \def\Amp{1/sqrt(\A)} 
    
    \draw [latex-latex] (0*\x,-3.5*\y) -- (0*\x,5*\y); 
    \draw [-latex] (-0.2*\x,0*\y) -- (2*\x,0*\y); 
    
    \draw[line width=2] plot[variable=\u, domain=0.24:2,samples=100] ({\u*\x},{(8*\B - 4/\u - 4*\A*\u)*\y}); 
    
    \draw [dashed] ({\um*\x},-2*\y) -- ({\um*\x},0.5*\y);

    \draw [dashed] ({\up*\x},-2*\y) -- ({\up*\x},0.5*\y);
    \draw [dashed] ({\Amp*\x},-0.3*\y) -- ({\Amp*\x},4*\y);
    
    \draw [decorate,decoration={brace,mirror,amplitude=6pt}] ({\um*\x},-2*\y) -- ({\up*\x},-2*\y);
    
    \node [] at ({\Amp*\x},-0.7*\y) {$\frac{1}{\sqrt{A}}$};
    \node [] at ({\um*\x+0.31},-0.3*\y) {$u_-$};
    \node [] at ({\up*\x-0.26},-0.3*\y) {$u_+$};
    \node [] at ({(\up+\um)*\x/2},-2.7*\y) {$\frac{2\sqrt{B^2-A}}{A}$};

    \node [] at (0.12*\x,4.6*\y) {$f(u)$};
    \node [] at (1.95*\x,0.3*\y) {$u$};
  \end{tikzpicture}
\end{center}
\caption{Plot of the function $f(u)$, showing two turning points $u_\pm$ that surround the region where $u'^2 > 0$, which is necessary for periodic solutions to exist. In this example $A>0$ and $B>\sqrt{A}$.}
\label{plotf}
\end{figure}
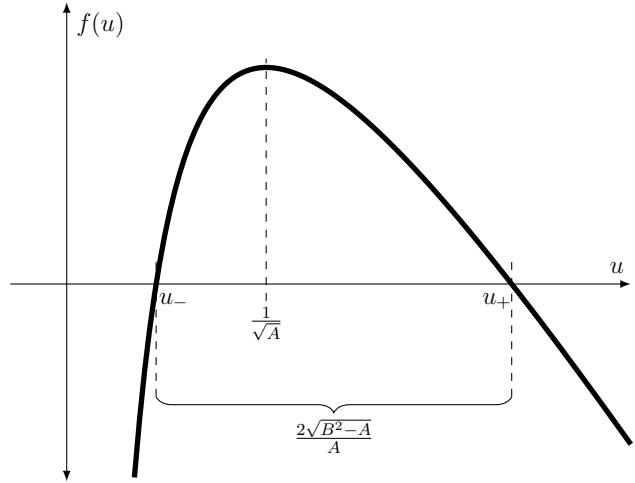

By first integrating the equation, then multiplying by the term  $\partial_\phi \left(\frac{1}{2\sqrt{\cL}}    \right)$, and then integrating once again, the equation readily becomes
\begin{equation}\label{eqint2}
  \sqrt{\cL}+\left[ \partial_\phi   \left(\frac{1}{2\sqrt{\cL}}    \right)   \right]^2= 2B-\frac{A}{\sqrt{\cL}}\, ,
\end{equation}
where $A$ and $B$ are integration constants. As a byproduct of such a procedure, the following identities are obtained
\begin{subequations}\label{conditions}
	\begin{align}
		\begin{split}\label{cond1}
				\left(\partial_\phi\LL\right)^2&=16\LL^{5/2}\left(2B\sqrt{\LL} -\LL-A\right)\,,
		\end{split}\\
		\begin{split}\label{cond2}
				\partial_\phi^2 \LL& = 4\LL^{3/2}\left(12B\sqrt{\LL}-7\LL-5A\right)\,,
		\end{split}
	\end{align}
\end{subequations}
which will prove to be useful later on. At this point, it is helpful to perform the field redefinition
\begin{equation}\label{Lu2}
  \cL=\frac{1}{u^2}\,,  
\end{equation}
by which Eq. \eqref{eqint2} becomes
\begin{equation}\label{up2}
 u'^2=f(u)\,,
\end{equation}
where the function $f(u)$ is given by
\begin{equation}\label{f}
 f(u)=-\frac{4A}{u}(u-u_-)(u-u_+),\quad u_\pm=\frac{B \pm \sqrt{B^2-A}}{A}\,.
\end{equation}
A plot of the function $f(u)$ can be found in Fig. \ref{plotf}.

Further integration of \eqref{up2} is straightforward by separation of variables
\begin{equation}\label{final_integration}
  \int \frac{du}{\pm\sqrt{8B-\frac{4}{u}-4Au}}=\phi  + \phi_0\,,
\end{equation}
rendering the third integration constant $\phi_0$, which corresponds to a phase. Although not particularly enlightening, the integral of the left-hand side can be evaluated explicitly, resulting in combinations of incomplete elliptic integrals of the first and second kind \cite{Mokhtari2011}, thus obscuring the possibility of inverting and expressing $u$ as a function of $\phi$. Furthermore, the explicit evaluation of the integral does not clarify whether the solution obeys periodic boundary conditions.

\begin{figure}[h]
  \centering
  \begin{subfigure}[b]{0.45\linewidth}
    \begin{center}
  \begin{tikzpicture}[scale=0.75, every node/.style={scale=0.8}]
    \def\x{1} 
    \def\y{2.2} 
    
    \def\A{2} 
    \def\B{1.8945} 
    
    \def\um{(-sqrt(\B^2-\A)+\B)/\A} 
    \def\up{(sqrt(\B^2-\A)+\B)/\A} 
    
    \def\Amp{1/sqrt(\A)} 

    \def\datau{(0.*\x , 0.316948*\y) (0.05*\x , 0.336338*\y) (0.1*\x , 0.389455*\y) (0.15*\x , 0.465676*\y) (0.2*\x , 0.555219*\y) (0.25*\x , 0.651181*\y) (0.3*\x , 0.749063*\y) (0.35*\x , 0.845935*\y) (0.4*\x , 0.939842*\y) (0.45*\x , 1.02944*\y) (0.5*\x , 1.11378*\y) (0.55*\x , 1.19217*\y) (0.6*\x , 1.26408*\y) (0.65*\x , 1.32913*\y) (0.7*\x , 1.38701*\y) (0.75*\x , 1.4375*\y) (0.8*\x , 1.48041*\y) (0.85*\x , 1.51561*\y) (0.9*\x , 1.54298*\y) (0.95*\x , 1.56246*\y) (1.*\x , 1.57399*\y) (1.05*\x , 1.57753*\y) (1.1*\x , 1.57309*\y) (1.15*\x , 1.56067*\y) (1.2*\x , 1.54031*\y) (1.25*\x , 1.51205*\y) (1.3*\x , 1.47598*\y) (1.35*\x , 1.43221*\y) (1.4*\x , 1.38089*\y) (1.45*\x , 1.32219*\y) (1.5*\x , 1.25635*\y) (1.55*\x , 1.18369*\y) (1.6*\x , 1.10461*\y) (1.65*\x , 1.01964*\y) (1.7*\x , 0.929506*\y) (1.75*\x , 0.83519*\y) (1.8*\x , 0.738094*\y) (1.85*\x , 0.640259*\y) (1.9*\x , 0.544755*\y) (1.95*\x , 0.456311*\y) (2.*\x , 0.382144*\y) (2.05*\x , 0.332313*\y) (2.1*\x , 0.317198*\y) (2.15*\x , 0.340792*\y) (2.2*\x , 0.397055*\y) (2.25*\x , 0.475206*\y) (2.3*\x , 0.56576*\y) (2.35*\x , 0.662126*\y) (2.4*\x , 0.760018*\y) (2.45*\x , 0.856641*\y) (2.5*\x , 0.950124*\y) (2.55*\x , 1.03918*\y) (2.6*\x , 1.12288*\y) (2.65*\x , 1.20056*\y) (2.7*\x , 1.27172*\y) (2.75*\x , 1.33598*\y) (2.8*\x , 1.39305*\y) (2.85*\x , 1.44269*\y) (2.9*\x , 1.48475*\y) (2.95*\x , 1.51907*\y) (3.*\x , 1.54556*\y) (3.05*\x , 1.56415*\y) (3.1*\x , 1.57478*\y) (3.15*\x , 1.57743*\y) (3.2*\x , 1.5721*\y) (3.25*\x , 1.55878*\y) (3.3*\x , 1.53753*\y) (3.35*\x , 1.50839*\y) (3.4*\x , 1.47146*\y) (3.45*\x , 1.42683*\y) (3.5*\x , 1.37467*\y) (3.55*\x , 1.31515*\y) (3.6*\x , 1.24854*\y) (3.65*\x , 1.17513*\y) (3.7*\x , 1.09536*\y) (3.75*\x , 1.00978*\y) (3.8*\x , 0.919117*\y) (3.85*\x , 0.82441*\y) (3.9*\x , 0.727115*\y) (3.95*\x , 0.629364*\y) (4.*\x , 0.534378*\y) (4.05*\x , 0.447121*\y) (4.1*\x , 0.375138*\y) (4.15*\x , 0.328732*\y) (4.2*\x , 0.317947*\y) (4.25*\x , 0.345662*\y) (4.3*\x , 0.404928*\y) (4.35*\x , 0.484889*\y) (4.4*\x , 0.576372*\y) (4.45*\x , 0.673089*\y) (4.5*\x , 0.770957*\y) (4.55*\x , 0.867308*\y) (4.6*\x , 0.96035*\y) (4.65*\x , 1.04884*\y) (4.7*\x , 1.1319*\y) (4.75*\x , 1.20887*\y) (4.8*\x , 1.27928*\y) (4.85*\x , 1.34274*\y) (4.9*\x , 1.39899*\y) (4.95*\x , 1.44779*\y) (5.*\x , 1.48898*\y) (5.05*\x , 1.52243*\y) (5.1*\x , 1.54804*\y) (5.15*\x , 1.56574*\y) (5.2*\x , 1.57548*\y) (5.25*\x , 1.57723*\y) (5.3*\x , 1.571*\y) (5.35*\x , 1.55679*\y) (5.4*\x , 1.53465*\y) (5.45*\x , 1.50464*\y) (5.5*\x , 1.46683*\y) (5.55*\x , 1.42136*\y) (5.6*\x , 1.36836*\y) (5.65*\x , 1.30803*\y) (5.7*\x , 1.24064*\y) (5.75*\x , 1.1665*\y) (5.8*\x , 1.08605*\y) (5.85*\x , 0.999848*\y) (5.9*\x , 0.908676*\y) (5.95*\x , 0.813597*\y) (6.*\x , 0.716128*\y) (6.05*\x , 0.618502*\y) (6.1*\x , 0.524096*\y) (6.15*\x , 0.438121*\y) (6.2*\x , 0.368453*\y) (6.25*\x , 0.325607*\y) (6.3*\x , 0.319192*\y) (6.35*\x , 0.35111*\y) (6.4*\x , 0.414796*\y) (6.45*\x , 0.501877*\y) }

    \draw [-latex] (0*\x,-0.2*\y) -- (0*\x,2.3*\y); 
    \draw [-latex] (-0.2*\x,0*\y) -- (2*3.14*\x+1,0*\y); 

    \draw [line width=2] plot [smooth] coordinates {\datau};
   
    \draw [dashed] (-0.2*\x,{\um*\y}) -- (2*3.14*\x+0.2,{\um*\y});
    \draw [dashed] (-0.2*\x,{\up*\y}) -- (2*3.14*\x+0.2,{\up*\y});
    \draw [dashed] (-0.2*\x,{\Amp*\y}) -- (2*3.14*\x+0.2,{\Amp*\y});

    \draw [dashed] (3.14/3*\x,{-0.1*\y}) -- (3.14/3*\x,{1.75*\y});
    \draw [dashed] (3.14*\x,{-0.1*\y}) -- (3.14*\x,{1.75*\y});
    \draw [dashed] (2*3.14*\x,{-0.1*\y}) -- (2*3.14*\x,{1.75*\y});

    \draw [decorate,decoration={brace,mirror,amplitude=6pt}] (3.14/3*\x,{-0.1*\y}) -- (3.14*\x,{-0.1*\y});
    \draw [decorate,decoration={brace,mirror,amplitude=6pt}] (2*3.14*\x+0.2,{\um*\y}) -- (2*3.14*\x+0.2,{\up*\y});

    \node [] at (2*3.14*\x+1.2,{(\up+\um)*\y/2}) {$\frac{2\sqrt{B^2-A}}{A}$};

    \node [] at (0.48*\x+0.1,2.2*\y) {$u(\phi)$};
    \node [] at (2*3.14*\x+0.8,0.15*\y) {$\phi$};

    \node [] at (2*3.14*\x,-0.15*\y) {$2\pi$};
    \node [] at (2*3.14/3*\x,-0.33*\y) {$T$};

    \node [] at (-0.5*\x,{\up*\y}) {$u_+$};
    \node [] at (-0.5*\x,{\um*\y}) {$u_-$};
    \node [] at (-0.5*\x,{\Amp*\y}) {$\frac{1}{\sqrt{A}}$};
  \end{tikzpicture}
\end{center}
\caption{Integration of the function $u(\phi)$.}
\label{plotu}
\end{subfigure}\hfill
\begin{subfigure}[b]{0.45\linewidth}
  \begin{center}
  \begin{tikzpicture}[scale=0.75, every node/.style={scale=0.8}]
    \def\x{1} 
    \def\y{0.4} 
    
    \def\A{2} 
    \def\B{1.8945} 
    
    \def\um{(-sqrt(\B^2-\A)+\B)/\A} 
    \def\up{(sqrt(\B^2-\A)+\B)/\A} 
    
    \def\Amp{1/sqrt(\A)} 

    \def\dataL{(0.*\x , 9.95461*\y) (0.05*\x , 8.83994*\y) (0.1*\x , 6.59302*\y) (0.15*\x , 4.61139*\y) (0.2*\x , 3.24393*\y) (0.25*\x , 2.35829*\y) (0.3*\x , 1.78223*\y) (0.35*\x , 1.39742*\y) (0.4*\x , 1.13211*\y) (0.45*\x , 0.943617*\y) (0.5*\x , 0.80612*\y) (0.55*\x , 0.703601*\y) (0.6*\x , 0.625823*\y) (0.65*\x , 0.566065*\y) (0.7*\x , 0.519803*\y) (0.75*\x , 0.483931*\y) (0.8*\x , 0.456283*\y) (0.85*\x , 0.435337*\y) (0.9*\x , 0.420027*\y) (0.95*\x , 0.409621*\y) (1.*\x , 0.403643*\y) (1.05*\x , 0.40183*\y) (1.1*\x , 0.404103*\y) (1.15*\x , 0.410561*\y) (1.2*\x , 0.421489*\y) (1.25*\x , 0.437389*\y) (1.3*\x , 0.459026*\y) (1.35*\x , 0.487511*\y) (1.4*\x , 0.524426*\y) (1.45*\x , 0.572025*\y) (1.5*\x , 0.633546*\y) (1.55*\x , 0.713713*\y) (1.6*\x , 0.819562*\y) (1.65*\x , 0.961843*\y) (1.7*\x , 1.15743*\y) (1.75*\x , 1.4336*\y) (1.8*\x , 1.83559*\y) (1.85*\x , 2.43943*\y) (1.9*\x , 3.36974*\y) (1.95*\x , 4.80263*\y) (2.*\x , 6.8477*\y) (2.05*\x , 9.05535*\y) (2.1*\x , 9.93892*\y) (2.15*\x , 8.61036*\y) (2.2*\x , 6.34305*\y) (2.25*\x , 4.42829*\y) (2.3*\x , 3.12418*\y) (2.35*\x , 2.28096*\y) (2.4*\x , 1.73122*\y) (2.45*\x , 1.36271*\y) (2.5*\x , 1.10774*\y) (2.55*\x , 0.926023*\y) (2.6*\x , 0.793112*\y) (2.65*\x , 0.693797*\y) (2.7*\x , 0.618325*\y) (2.75*\x , 0.560274*\y) (2.8*\x , 0.515311*\y) (2.85*\x , 0.480454*\y) (2.9*\x , 0.453624*\y) (2.95*\x , 0.433356*\y) (3.*\x , 0.418627*\y) (3.05*\x , 0.408737*\y) (3.1*\x , 0.403236*\y) (3.15*\x , 0.401882*\y) (3.2*\x , 0.404615*\y) (3.25*\x , 0.411556*\y) (3.3*\x , 0.423013*\y) (3.35*\x , 0.439513*\y) (3.4*\x , 0.461855*\y) (3.45*\x , 0.491195*\y) (3.5*\x , 0.529181*\y) (3.55*\x , 0.578159*\y) (3.6*\x , 0.641502*\y) (3.65*\x , 0.724144*\y) (3.7*\x , 0.833456*\y) (3.75*\x , 0.980728*\y) (3.8*\x , 1.18375*\y) (3.85*\x , 1.47134*\y) (3.9*\x , 1.89145*\y) (3.95*\x , 2.52462*\y) (4.*\x , 3.50189*\y) (4.05*\x , 5.00206*\y) (4.1*\x , 7.10587*\y) (4.15*\x , 9.25372*\y) (4.2*\x , 9.89216*\y) (4.25*\x , 8.36944*\y) (4.3*\x , 6.0988*\y) (4.35*\x , 4.2532*\y) (4.4*\x , 3.01019*\y) (4.45*\x , 2.20727*\y) (4.5*\x , 1.68244*\y) (4.55*\x , 1.32939*\y) (4.6*\x , 1.08428*\y) (4.65*\x , 0.909033*\y) (4.7*\x , 0.780521*\y) (4.75*\x , 0.684289*\y) (4.8*\x , 0.611043*\y) (4.85*\x , 0.554645*\y) (4.9*\x , 0.510944*\y) (4.95*\x , 0.477078*\y) (5.*\x , 0.451047*\y) (5.05*\x , 0.431444*\y) (5.1*\x , 0.417288*\y) (5.15*\x , 0.407908*\y) (5.2*\x , 0.40288*\y) (5.25*\x , 0.401984*\y) (5.3*\x , 0.40518*\y) (5.35*\x , 0.412608*\y) (5.4*\x , 0.4246*\y) (5.45*\x , 0.441709*\y) (5.5*\x , 0.464771*\y) (5.55*\x , 0.494988*\y) (5.6*\x , 0.534074*\y) (5.65*\x , 0.584471*\y) (5.7*\x , 0.649697*\y) (5.75*\x , 0.734906*\y) (5.8*\x , 0.847821*\y) (5.85*\x , 1.0003*\y) (5.9*\x , 1.2111*\y) (5.95*\x , 1.51071*\y) (6.*\x , 1.94993*\y) (6.05*\x , 2.61407*\y) (6.1*\x , 3.64065*\y) (6.15*\x , 5.2097*\y) (6.2*\x , 7.36607*\y) (6.25*\x , 9.43221*\y) (6.3*\x , 9.81513*\y) (6.35*\x , 8.11173*\y) (6.4*\x , 5.81206*\y) (6.45*\x , 3.97013*\y) }

    \draw [-latex] (0*\x,-0.2*\y) -- (0*\x,13*\y); 
    \draw [-latex] (-0.2*\x,0*\y) -- (2*3.14*\x+1,0*\y); 

    \draw [line width=2] plot [smooth] coordinates {\dataL};
   
    \draw [dashed] (-0.2*\x,{1/(\um)^2*\y}) -- (2*3.14*\x+0.2,{1/(\um)^2*\y});
    \draw [dashed] (-0.2*\x,{1/(\up)^2*\y}) -- (2*3.14*\x+0.2,{1/(\up)^2*\y});
    \draw [dashed] (-0.2*\x,{\A*\y}) -- (2*3.14*\x+0.2,{\A*\y});

    \draw [dashed] (3.14/3*\x,{-0.1*\y}) -- (3.14/3*\x,{1.2*\y});
    \draw [dashed] (3.14*\x,{-0.1*\y}) -- (3.14*\x,{1.2*\y});
    \draw [dashed] (2*3.14*\x,{-0.1*\y}) -- (2*3.14*\x,{11*\y});

    \draw [decorate,decoration={brace,mirror,amplitude=6pt}] (3.14/3*\x,{-0.4*\y}) -- (3.14*\x,{-0.4*\y});
    \draw [decorate,decoration={brace,amplitude=6pt}] (2*3.14*\x+0.3,{1/(\um)^2*\y}) -- (2*3.14*\x+0.3,{1/(\up)^2*\y});

    \node [] at (2*3.14*\x+1.5,{(1/(\up)^2+1/(\um)^2)*\y/2}) {{\scriptsize $4B\sqrt{B^2-A}$}};

    \node [] at (0.55*\x,12*\y) {$\cL (\phi)$};
    \node [] at (2*3.14*\x+0.8,0.8*\y) {$\phi$};

    \node [] at (2*3.14*\x,-0.6*\y) {$2\pi$};
    \node [] at (2*3.14/3*\x,-1.59*\y) {$T$};

    \node [] at (-0.7*\x,{1/(\up)^2*\y}) {$1/u_+^2$};
    \node [] at (-0.7*\x,{1/(\um)^2*\y}) {$1/u_-^2$};
    \node [] at (-0.5*\x,{\A*\y}) {$A$};
 
  \end{tikzpicture}
\end{center}
\caption{Integration of the function $\cL(\phi)$.}
\label{plotL}
\end{subfigure}
\caption{Numerical integration of equation \eqref{up2}. In this example, $A=2$ and $B=1.8945$, thus rendering a periodic solution with $T=\frac{2\pi}{3}$.}
\label{plotuL}
\end{figure}
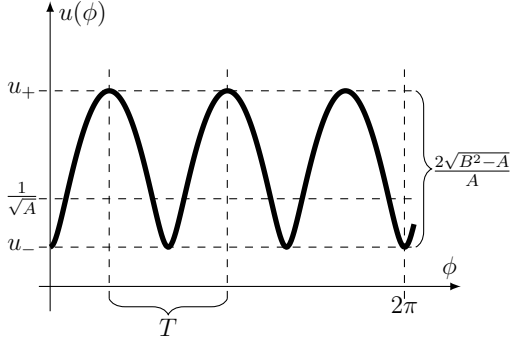
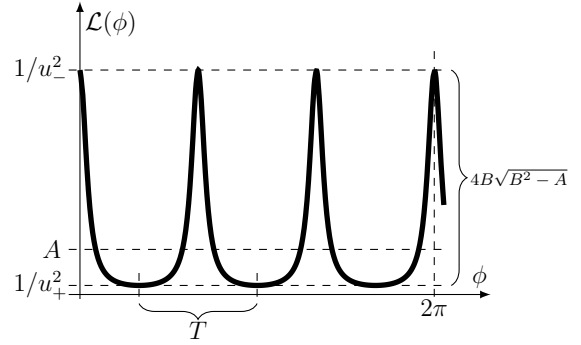

The necessary condition to obtain real periodic solutions can be better understood from Eq. \eqref{up2}: the function $f(u)$ should be positive in a range between two real positive zeros $u_\pm$, assuring that the function $u(\phi)$ oscillates between two positive turning points. These conditions are met by imposing $A>0$ and $B>\sqrt{A}$ on the integration constants (see Fig. \ref{plotf}). Once these conditions are fulfilled, a simple numerical integration of \eqref{up2} renders the periodic solution for $u(\phi)$ shown in Fig. \ref{plotu}. The function $\cL(\phi)$ can then be obtained from the relation \eqref{Lu2} (see Fig. \ref{plotL}).

Intricate relations among the parameters are a staple characteristic of periodic solutions in nonlinear integrable equations, and the solution presented here is not an exception. For example, the inflection point located at $u=\frac{1}{\sqrt{A}}$ gives the characteristic ``height'' of the oscillation. On the other hand,  the ``amplitude,'' i.e., the difference between the turning points, is given by $u_+-u_-=\frac{2 \sqrt{B^2-A}}{A}$. Moreover, the period $T$ of the solution can be obtained from \eqref{final_integration}, yielding
\begin{equation}\label{period}
  T=2\int_{u_-}^{u_+} \frac{du}{\sqrt{8B-\frac{4}{u}-4Au}}\,.
\end{equation}

\begin{figure}[h]
  \centering
  \begin{subfigure}[b]{0.45\linewidth}
    \begin{center}
      \begin{tikzpicture}[xscale=1,yscale=0.4,every node/.style={scale=0.85}]
        
        \def\dataa{(0. , 0.016303) (0.1 , 1.47856) (0.2 , 2.89223) (0.3 , 4.20862) (0.4 , 5.42621) (0.5 , 6.54449) (0.6 , 7.56325) (0.7 , 8.48236) (0.8 , 9.30175) (0.9 , 10.0214) (1. , 10.6412) (1.1 , 11.1612) (1.2 , 11.5813) (1.3 , 11.9017) (1.4 , 12.1221) (1.5 , 12.2427) (1.6 , 12.2634) (1.7 , 12.1843) (1.8 , 12.0053) (1.9 , 11.7264) (2. , 11.3477) (2.1 , 10.8691) (2.2 , 10.2907) (2.3 , 9.61248) (2.4 , 8.83449) (2.5 , 7.95675) (2.6 , 6.97933) (2.7 , 5.90233) (2.8 , 4.72592) (2.9 , 3.45043) (3. , 2.07678) (3.1 , 0.609513) (3.2 , 0.864852) (3.3 , 2.31722) (3.4 , 3.6743) (3.5 , 4.93297) (3.6 , 6.0925) (3.7 , 7.15257) (3.8 , 8.11304) (3.9 , 8.97382) (4. , 9.73484) (4.1 , 10.3961) (4.2 , 10.9575) (4.3 , 11.4191) (4.4 , 11.7808) (4.5 , 12.0427) (4.6 , 12.2048) (4.7 , 12.2669) (4.8 , 12.2292) (4.9 , 12.0916) (5. , 11.8542) (5.1 , 11.5169) (5.2 , 11.0798) (5.3 , 10.5428) (5.4 , 9.906) (5.5 , 9.1694) (5.6 , 8.33305) (5.7 , 7.39699) (5.8 , 6.36129) (5.9 , 5.2261) (6. , 3.99167) (6.1 , 2.65856) (6.2 , 1.22879) (6.28 , 0.016303) }
        
        \def\dataaa{ (0. , 0.0373403) (0.1 , 0.922976) (0.2 , 1.8026) (0.3 , 2.58924) (0.4 , 3.27902) (0.5 , 3.8707) (0.6 , 4.36374) (0.7 , 4.75783) (0.8 , 5.05282) (0.9 , 5.24859) (1. , 5.3451) (1.1 , 5.3423) (1.2 , 5.24021) (1.3 , 5.03884) (1.4 , 4.73827) (1.5 , 4.3386) (1.6 , 3.84) (1.7 , 3.24277) (1.8 , 2.5475) (1.9 , 1.75547) (2. , 0.870913) (2.1 , 0.0566607) (2.2 , 0.974796) (2.3 , 1.84943) (2.4 , 2.63067) (2.5 , 3.31495) (2.6 , 3.90109) (2.7 , 4.38856) (2.8 , 4.77708) (2.9 , 5.06648) (3. , 5.25666) (3.1 , 5.34757) (3.2 , 5.33918) (3.3 , 5.2315) (3.4 , 5.02455) (3.5 , 4.71839) (3.6 , 4.31314) (3.7 , 3.80897) (3.8 , 3.20621) (3.9 , 2.50544) (4. , 1.70803) (4.1 , 0.818608) (4.2 , 0.0973217) (4.3 , 1.02637) (4.4 , 1.89597) (4.5 , 2.67181) (4.6 , 3.35058) (4.7 , 3.93117) (4.8 , 4.41307) (4.9 , 4.79601) (5. , 5.07983) (5.1 , 5.26442) (5.2 , 5.34974) (5.3 , 5.33575) (5.4 , 5.22248) (5.5 , 5.00993) (5.6 , 4.69819) (5.7 , 4.28736) (5.8 , 3.77763) (5.9 , 3.16933) (6. , 2.46307) (6.1 , 1.6603) (6.2 , 0.766074) (6.28 , 0.0373403)}
        
        \def\dataaaa{(0. , 0.0692244) (0.1 , 0.629771) (0.2 , 1.22951) (0.3 , 1.74434) (0.4 , 2.16607) (0.5 , 2.49215) (0.6 , 2.7215) (0.7 , 2.85357) (0.8 , 2.88811) (0.9 , 2.82506) (1. , 2.66454) (1.1 , 2.40686) (1.2 , 2.05268) (1.3 , 1.60339) (1.4 , 1.06236) (1.5 , 0.44293) (1.6 , 0.182615) (1.7 , 0.812584) (1.8 , 1.38927) (1.9 , 1.87733) (2. , 2.27128) (2.1 , 2.56919) (2.2 , 2.77016) (2.3 , 2.87375) (2.4 , 2.87978) (2.5 , 2.78824) (2.6 , 2.59929) (2.7 , 2.31332) (2.8 , 1.93117) (2.9 , 1.45456) (3. , 0.888214) (3.1 , 0.257135) (3.2 , 0.363652) (3.3 , 0.989517) (3.4 , 1.54138) (3.5 , 2.00227) (3.6 , 2.36829) (3.7 , 2.63795) (3.8 , 2.8105) (3.9 , 2.8856) (4. , 2.86312) (4.1 , 2.7431) (4.2 , 2.52575) (4.3 , 2.21157) (4.4 , 1.80158) (4.5 , 1.29801) (4.6 , 0.707737) (4.7 , 0.0966781) (4.8 , 0.551053) (4.9 , 1.15969) (5. , 1.68568) (5.1 , 2.1191) (5.2 , 2.45708) (5.3 , 2.69842) (5.4 , 2.84253) (5.5 , 2.88913) (5.6 , 2.83814) (5.7 , 2.68964) (5.8 , 2.44394) (5.9 , 2.10164) (6. , 1.66398) (6.1 , 1.13395) (6.2 , 0.522256) (6.28 , 0.0692244)}
        
        \def\dataaaaa{(0. , 0.117889) (0.1 , 0.448561) (0.2 , 0.858635) (0.3 , 1.1995) (0.4 , 1.45492) (0.5 , 1.61999) (0.6 , 1.69277) (0.7 , 1.67261) (0.8 , 1.55966) (0.9 , 1.35506) (1. , 1.06168) (1.1 , 0.687284) (1.2 , 0.266245) (1.3 , 0.215264) (1.4 , 0.63255) (1.5 , 1.01639) (1.6 , 1.32116) (1.7 , 1.53775) (1.8 , 1.66295) (1.9 , 1.69545) (2. , 1.63498) (2.1 , 1.48207) (2.2 , 1.23846) (2.3 , 0.908423) (2.4 , 0.50539) (2.5 , 0.129279) (2.6 , 0.391735) (2.7 , 0.807564) (2.8 , 1.15903) (2.9 , 1.42617) (3. , 1.60337) (3.1 , 1.68846) (3.2 , 1.68063) (3.3 , 1.57995) (3.4 , 1.3874) (3.5 , 1.10552) (3.6 , 0.741004) (3.7 , 0.320518) (3.8 , 0.17028) (3.9 , 0.576941) (4. , 0.969686) (4.1 , 1.28569) (4.2 , 1.51423) (4.3 , 1.65165) (4.4 , 1.69649) (4.5 , 1.64834) (4.6 , 1.50763) (4.7 , 1.27588) (4.8 , 0.956877) (4.9 , 0.561844) (5. , 0.159733) (5.1 , 0.335468) (5.2 , 0.755277) (5.3 , 1.11706) (5.4 , 1.39583) (5.5 , 1.58514) (5.6 , 1.68251) (5.7 , 1.68702) (5.8 , 1.59862) (5.9 , 1.41816) (6. , 1.14788) (6.1 , 0.793605) (6.2 , 0.376504) (6.28 , 0.117889)}
        
        \def\dataaaaaa{(0. , 0.204948) (0.1 , 0.34797) (0.2 , 0.582894) (0.3 , 0.782384) (0.4 , 0.91586) (0.5 , 0.973659) (0.6 , 0.95285) (0.7 , 0.854424) (0.8 , 0.684165) (0.9 , 0.459319) (1. , 0.243701) (1.1 , 0.252618) (1.2 , 0.47272) (1.3 , 0.695409) (1.4 , 0.861929) (1.5 , 0.956098) (1.6 , 0.97249) (1.7 , 0.910328) (1.8 , 0.772803) (1.9 , 0.570172) (2. , 0.335339) (2.1 , 0.205539) (2.2 , 0.360805) (2.3 , 0.595486) (2.4 , 0.791754) (2.5 , 0.921152) (2.6 , 0.97458) (2.7 , 0.949358) (2.8 , 0.846691) (2.9 , 0.67274) (3. , 0.445901) (3.1 , 0.235529) (3.2 , 0.262206) (3.3 , 0.486089) (3.4 , 0.706469) (3.5 , 0.869204) (3.6 , 0.959099) (3.7 , 0.971073) (3.8 , 0.904559) (3.9 , 0.763012) (4. , 0.55733) (4.1 , 0.322954) (4.2 , 0.207302) (4.3 , 0.37381) (4.4 , 0.607942) (4.5 , 0.80091) (4.6 , 0.926203) (4.7 , 0.975252) (4.8 , 0.945621) (4.9 , 0.838732) (5. , 0.66114) (5.1 , 0.432485) (5.2 , 0.228176) (5.3 , 0.272394) (5.4 , 0.499409) (5.5 , 0.717341) (5.6 , 0.876247) (5.7 , 0.961855) (5.8 , 0.969408) (5.9 , 0.898551) (6. , 0.753015) (6.1 , 0.544375) (6.2 , 0.310857) (6.28 , 0.204948)}

        \draw [-latex] (0,-1) -- (0,12); 
        \draw [-latex] (-0.25,-0.5) -- (2*3.14+0.5,-0.5); 

        \draw [line width=0.5] plot [smooth] coordinates {\dataa};
        \draw [line width=0.5] plot [smooth] coordinates {\dataaa};
        \draw [line width=0.5] plot [smooth] coordinates {\dataaaa};
        \draw [line width=0.5] plot [smooth] coordinates {\dataaaaa};
        \draw [line width=0.5] plot [smooth] coordinates {\dataaaaaa};

        \draw [dashed] (-0.1,{1/sqrt(5)}) -- (2*3.14+0.1,{1/sqrt(5)});

        \node [] at (-0.4,{1/sqrt(5)}) {$\frac{1}{\sqrt{A}}$};
        
        \node [] at (0.4,11.5) {$u (\phi)$};
        \node [] at (2*3.14+0.4,0.3) {$\phi$};
        
      \end{tikzpicture}
    \end{center}
    \caption{Plots of the function $u(\phi)$ for $A=5$ and various mode values $ m = \{2, 3, 4, 5, 6\}$\vspace{11pt}  .
 }
    \label{plotuvariosn}
  \end{subfigure}\hfill
  \begin{subfigure}[b]{0.45\linewidth}
    \begin{center}
      \begin{tikzpicture}[xscale=0.35*1.2,yscale=0.75*1.2, every node/.style={scale=0.6}]
        
        \def\x{1}
        \def\y{1}
        
        \def\data{(0.3996, 0.64)  (0.4329, 0.78)  (0.4554, 0.88)  (0.4831, 1.01)  (0.5036, 1.11)  (0.5249, 1.22)  (0.5439, 1.32)  (0.5634, 1.43)  (0.5809, 1.53)  (0.5984, 1.63)  (0.6161, 1.74)  (0.6326, 1.84)  (0.6481, 1.94)  (0.6636, 2.04)  (0.6829, 2.17)  (0.7044, 2.32)  (0.723, 2.45)  (0.7421, 2.59)  (0.7609, 2.73)  (0.7789, 2.87)  (0.7955, 3.)  (0.8145, 3.15)  (0.8304, 3.28)  (0.8474, 3.42)  (0.8649, 3.57)  (0.88, 3.7)  (0.8981, 3.86)  (0.9126, 3.99)  (0.9245, 4.1)  (0.9374, 4.22)  (0.9494, 4.33)  (0.9611, 4.44)  (0.9711, 4.54)  (0.9816, 4.64)  (0.993, 4.75)  (1.0099, 4.92)}
        
        \def\datb{(1.0041, 1.01)  (1.0431, 1.11)  (1.0844, 1.22)  (1.1209, 1.32)  (1.1594, 1.43)  (1.1934, 1.53)  (1.2264, 1.63)  (1.2621, 1.74)  (1.2931, 1.84)  (1.3236, 1.94)  (1.3536, 2.04)  (1.3919, 2.17)  (1.4344, 2.32)  (1.47, 2.45)  (1.5081, 2.59)  (1.5449, 2.73)  (1.5809, 2.87)  (1.6135, 3.)  (1.6505, 3.15)  (1.6814, 3.28)  (1.7149, 3.42)  (1.7494, 3.57)  (1.779, 3.7)  (1.8151, 3.86)  (1.8436, 3.99)  (1.8675, 4.1)  (1.8929, 4.22)  (1.9164, 4.33)  (1.9391, 4.44)  (1.9601, 4.54)  (1.9801, 4.64)  (2.0025, 4.75)  (2.0364, 4.92)}

        \def\datc{(1.7219, 1.32)  (1.7789, 1.43)  (1.8284, 1.53)  (1.8774, 1.63)  (1.9291, 1.74)  (1.9751, 1.84)  (2.0206, 1.94)  (2.0646, 2.04)  (2.1204, 2.17)  (2.1834, 2.32)  (2.2365, 2.45)  (2.2921, 2.59)  (2.3464, 2.73)  (2.3999, 2.87)  (2.448, 3.)  (2.5025, 3.15)  (2.5494, 3.28)  (2.5984, 3.42)  (2.6499, 3.57)  (2.694, 3.7)  (2.7471, 3.86)  (2.7896, 3.99)  (2.825, 4.1)  (2.8629, 4.22)  (2.8974, 4.33)  (2.9316, 4.44)  (2.9621, 4.54)  (2.9926, 4.64)  (3.0255, 4.75)  (3.0759, 4.92)}
        
        \def\datd{(2.5216, 1.59)  (2.5905, 1.7)  (2.6515, 1.8)  (2.711, 1.9)  (2.7695, 2.)  (2.8379, 2.12)  (2.9105, 2.25)  (2.992, 2.4)  (3.0604, 2.53)  (3.1329, 2.67)  (3.2089, 2.82)  (3.273, 2.95)  (3.3411, 3.09)  (3.4074, 3.23)  (3.4724, 3.37)  (3.532, 3.5)  (3.5995, 3.65)  (3.6569, 3.78)  (3.7266, 3.94)  (3.7781, 4.06)  (3.8244, 4.17)  (3.8746, 4.29)  (3.92, 4.4)  (3.9605, 4.5)  (4.001, 4.6)  (4.0446, 4.71)  (4.1119, 4.88)  (4.1585, 5.)}
        
        \def\datf{(3.399, 1.85)  (3.4715, 1.95)  (3.543, 2.05)  (3.6406, 2.19)  (3.7354, 2.33)  (3.8216, 2.46)  (3.9181, 2.61)  (4.0006, 2.74)  (4.0869, 2.88)  (4.1719, 3.02)  (4.2551, 3.16)  (4.3311, 3.29)  (4.4166, 3.44) (4.4954, 3.58)  (4.5671, 3.71)  (4.6544, 3.87)  (4.7235, 4.)  (4.7869, 4.12)  (4.8439, 4.23)  (4.9006, 4.34)  (4.9565, 4.45)  (5.007, 4.55)  (5.057, 4.65)  (5.1305, 4.8)  (5.1939, 4.93)}
        
        \def\datg{(4.3351, 2.09)  (4.4499, 2.23)  (4.5614, 2.37)  (4.663, 2.5)  (4.7775, 2.65)  (4.8744, 2.78)  (4.9769, 2.92)  (5.0844, 3.07)  (5.1755, 3.2)  (5.2721, 3.34)  (5.3669, 3.48)  (5.4604, 3.62)  (5.5455, 3.75)  (5.6486, 3.91)  (5.7306, 4.04)  (5.7995, 4.15)  (5.8676, 4.26)  (5.9409, 4.38)  (6.0014, 4.48)  (6.0614, 4.58)  (6.1261, 4.69)  (6.2084, 4.83)  (6.2949, 4.98)}
        
        \def\dath{(5.3274, 2.32)  (5.446, 2.45)  (5.5706, 2.59)  (5.6924, 2.73)  (5.8114, 2.87)  (5.92, 3.)  (6.043, 3.15)  (6.1479, 3.28)  (6.2584, 3.42)  (6.3749, 3.57)  (6.474, 3.7)  (6.5941, 3.86)  (6.6901, 3.99)  (6.7705, 4.1)  (6.8569, 4.22)  (6.9349, 4.33)  (7.0126, 4.44)  (7.0821, 4.54)  (7.1511, 4.64)  (7.226, 4.75)  (7.3409, 4.92)}
        
        \def\dati{(6.3599, 2.53)  (6.4994, 2.67)  (6.6454, 2.82)  (6.7695, 2.95)  (6.9006, 3.09)  (7.0289, 3.23)  (7.1554, 3.37)  (7.271, 3.5)  (7.402, 3.65)  (7.5134, 3.78)  (7.6486, 3.94)  (7.7486, 4.06)  (7.8389, 4.17)  (7.9366, 4.29)  (8.0245, 4.4)  (8.104, 4.5)  (8.183, 4.6)  (8.2686, 4.71)  (8.3994, 4.88)  (8.4905, 5.)}
        
        \def\datj{(7.4339, 2.73)  (7.5849, 2.87)  (7.7225, 3.)  (7.8785, 3.15)  (8.0109, 3.28)  (8.1514, 3.42)  (8.2989, 3.57)  (8.425, 3.7)  (8.5776, 3.86)  (8.6996, 3.99)  (8.8015, 4.1)  (8.9114, 4.22)  (9.0109, 4.33)  (9.1091, 4.44)  (9.1976, 4.54)  (9.2856, 4.64)  (9.381, 4.75)  (9.5269, 4.92)}
        
        \def\datk{(8.5686, 2.94)  (8.7304, 3.08)  (8.8781, 3.21)  (9.0456, 3.36)  (9.1881, 3.49)  (9.3394, 3.63)  (9.4879, 3.77)  (9.6449, 3.92)  (9.779, 4.05)  (9.8906, 4.16)  (10.0114, 4.28)  (10.1206, 4.39)  (10.2191, 4.49)  (10.3166, 4.59)  (10.4225, 4.7)  (10.5561, 4.84)  (10.6976, 4.99)}
        
        \def\datl{(9.7134, 3.12)  (9.8745, 3.25)  (10.0565, 3.4)  (10.2114, 3.53)  (10.3764, 3.67)  (10.5499, 3.82)  (10.7091, 3.96)  (10.8439, 4.08)  (10.977, 4.2)  (11.0976, 4.31)  (11.2169, 4.42)  (11.3244, 4.52)  (11.4309, 4.62)  (11.5469, 4.73)  (11.7235, 4.9)}
        
        \def\datm{(10.9264, 3.32)  (11.0965, 3.45)  (11.2766, 3.59)  (11.4544, 3.73)  (11.6414, 3.88)  (11.8011, 4.01)  (11.9464, 4.13)  (12.0786, 4.24)  (12.2206, 4.36)  (12.3381, 4.46)  (12.4541, 4.56)  (12.5696, 4.66)  (12.7401, 4.81)  (12.9086, 4.96)}
        
        \def\datn{(12.1406, 3.49)  (12.3339, 3.63)  (12.5244, 3.77)  (12.7249, 3.92)  (12.8965, 4.05)  (13.0401, 4.16)  (13.1944, 4.28)  (13.3346, 4.39)  (13.4611, 4.49)  (13.5861, 4.59)  (13.722, 4.7)  (13.8936, 4.84)  (14.0751, 4.99)}
        
        \def\dato{(13.3916, 3.66)  (13.5806, 3.79)  (13.8095, 3.95)  (13.9789, 4.07)  (14.1324, 4.18)  (14.2975, 4.3)  (14.4476, 4.41)  (14.5826, 4.51)  (14.7166, 4.61)  (14.8624, 4.72)  (15.0851, 4.89)}
        
        \def\datp{(14.7241, 3.86)  (14.9216, 3.99)  (15.0865, 4.1)  (15.2644, 4.22)  (15.4259, 4.33)  (15.5856, 4.44)  (15.7291, 4.54)  (15.8716, 4.64)  (16.027, 4.75)  (16.2639, 4.92)}

        \draw [-latex] (0*\x,-0.2*0.35/0.75*\y) -- (0*\x,5.5*\y); 
        \draw [-latex] (-0.2*\x,0*\y) -- (16.8*\x,0*\y); 
        
        \foreach \x in {1,2,3,4,5,6,7,8,9,10,11,12,13,14,15,16}
        \draw (\x cm,1pt) -- (\x cm,-1pt) node[anchor=north] {{\tiny$\x$}};
        \foreach \y in {1,2,3,4,5}
        \draw (1pt,\y cm) -- (-1pt,\y cm) node[anchor=east] {{\tiny$\y$}};
        
        \draw [] plot [smooth] coordinates {\data};
        \draw [] plot [smooth] coordinates {\datb};
        \draw [] plot [smooth] coordinates {\datc};
        \draw [] plot [smooth] coordinates {\datd};
        \draw [] plot [smooth] coordinates {\datf};
        \draw [] plot [smooth] coordinates {\datg};
        \draw [] plot [smooth] coordinates {\dath};
        \draw [] plot [smooth] coordinates {\dati};
        \draw [] plot [smooth] coordinates {\datj};
        \draw [] plot [smooth] coordinates {\datk};
        \draw [] plot [smooth] coordinates {\datl};
        \draw [] plot [smooth] coordinates {\datm};
        \draw [] plot [smooth] coordinates {\datn};
        \draw [] plot [smooth] coordinates {\dato};
        \draw [] plot [smooth] coordinates {\datp};
        
        \draw[densely dotted,line width=1] plot[variable=\B, domain=0:4,samples=20] ({\B^2},{\B}); 
        \fill[lightgray, variable=\B, domain=0:4,samples=20] (0,0) -- plot({\B^2},{\B}) -- (16,0); 

        \node [] at (0.4*\x,5.3*\y) {$B$};
        \node [] at (16.5*\x,0.3*\y) {$A$};
        
        \node [rotate=87] at (0.76*\x,4.5*\y) {{\tiny$m=1$}};
        \node [rotate=84] at (1.75*\x,4.5*\y) {{\tiny$m=2$}};
        \node [rotate=81] at (2.75*\x,4.5*\y) {{\tiny$m=3$}};
        \node [rotate=79] at (3.76*\x,4.5*\y) {{\tiny$m=4$}};
        \node [rotate=76] at (4.77*\x,4.5*\y) {{\tiny$m=5$}};
        \node [rotate=74] at (5.81*\x,4.5*\y) {{\tiny$m=6$}};
        \node [rotate=72] at (6.85*\x,4.5*\y) {{\tiny$m=7$}};
        \node [rotate=69.5] at (7.9*\x,4.5*\y) {{\tiny$m=8$}};
        \node [rotate=67] at (8.95*\x,4.5*\y) {{\tiny$m=9$}};
        \node [rotate=65] at (10.02*\x,4.5*\y) {{\tiny$m=10$}};
        \node [rotate=62.5] at (11.1*\x,4.5*\y) {{\tiny$m=11$}};
        \node [rotate=61.5] at (12.17*\x,4.5*\y) {{\tiny$m=12$}};
        \node [rotate=59.5] at (13.24*\x,4.5*\y) {{\tiny$m=13$}};
        \node [rotate=57] at (14.35*\x,4.5*\y) {{\tiny$m=14$}};
        \node [rotate=56] at (15.45*\x,4.5*\y) {{\tiny$m=15$}};

        \node[fill=white,draw,circle,inner sep=2pt]at (0.40,0.63){};
        \node[fill=white,draw,circle,inner sep=2pt]at (1,1){}; 
        \node[fill=white,draw,circle,inner sep=2pt]at (1.72,1.31){};
        \node[fill=white,draw,circle,inner sep=2pt]at (2.52,1.59){};
        \node[fill=white,draw,circle,inner sep=2pt]at (3.39,1.84){};
        \node[fill=white,draw,circle,inner sep=2pt]at (4.33,2.08){};
        \node[fill=white,draw,circle,inner sep=2pt]at (5.31,2.31){};
        \node[fill=white,draw,circle,inner sep=2pt]at (6.35,2.52){};
        \node[fill=white,draw,circle,inner sep=2pt]at (7.43,2.73){};
        \node[fill=white,draw,circle,inner sep=2pt]at (8.55,2.92){};
        \node[fill=white,draw,circle,inner sep=2pt]at (9.71,3.12){};
        \node[fill=white,draw,circle,inner sep=2pt]at (10.90,3.30){};
        \node[fill=white,draw,circle,inner sep=2pt]at (12.13,3.48){};
        \node[fill=white,draw,circle,inner sep=2pt]at (13.39,3.66){};
        \node[fill=white,draw,circle,inner sep=2pt]at (14.68,3.83){};

      \end{tikzpicture}
    \end{center}
    \caption{Parameter space for stationary periodic solutions. Values for $A$ and $B$ within the shaded area, below the dotted parabola $A = B^2$, are not allowed. The loci where the period corresponds to $2\pi/m$ are shown.
} 
    \label{AB}
  \end{subfigure}
  \caption{Exploring the parameter space of stationary periodic solutions of the Dym equation. As can be seen in the plot on the right, for $A=5$, the maximum attainable mode is $m=6$, which is the highest displayed in the plot on the left. The limiting values $A_{\min}^{(m)}$ and $B_{\min}^{(m)}$ delimiting the parameter region supporting the $m$th mode (represented by small white circles in the plot on the right) are given by
$\left(B_{\min}^{(m)}\right)^2=A_{\min}^{(m)}=\left(\frac{m}{2}\right)^{4/3}$, which follows by evaluating the period integral \eqref{period} in the limit $B\rightarrow\sqrt{A}$.}
  \label{figura3} 
\end{figure}
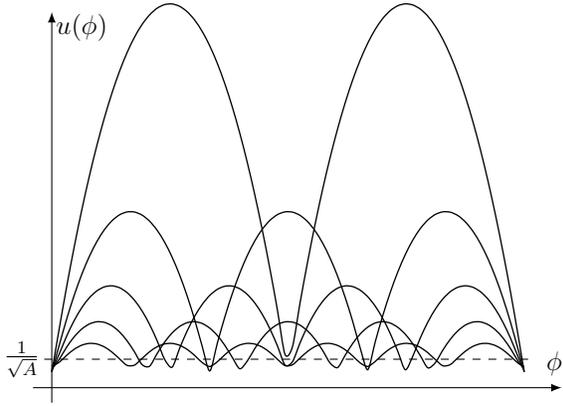
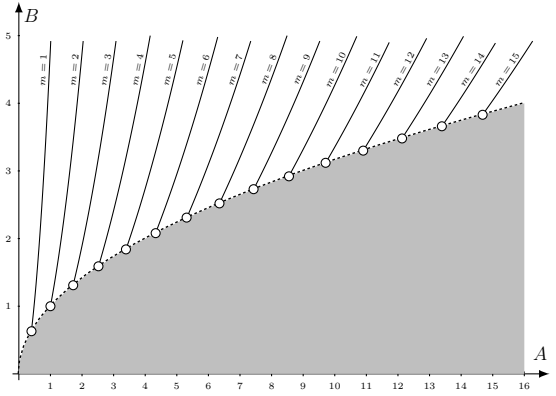

Since we are interested in periodic solutions in the range $0<\phi<2\pi$, the period is restricted to $T=\frac{2\pi}{m}$, where $m$ is a positive integer representing the ``mode'' of the solution. Hence, for a chosen value of $m$, the parameters $A$ and $B$ are related to each other since the integral in \eqref{period} must evaluate to a period $2\pi/m$. Hence, the constant $B$ can be obtained from \eqref{period} once $A$ and $m$ are chosen (see Fig. \ref{AB}). 

Although not evident from the relation \eqref{period}, for a fixed value of $A$, both the magnitude of $B$ and the amplitude decrease as the mode $m$ increases, as illustrated in Fig. \ref{plotuvariosn}. However, arbitrarily large values for $m$ are not supported for a given $A$, as can be seen in Fig. \ref{AB}: for a given value of $A$ (or $B$), only a finite number of modes can be reached. Equivalently, attaining a periodic solution with a given mode $m$ requires the parameters $A$ and $B$ to exceed corresponding threshold values, as shown in Fig. \ref{AB}.

After finding the periodic solutions, we can evaluate the Dym conserved charges $H_n$ shown in Appendix \ref{cargasDym}. For this purpose, relations \eqref{cond1} and \eqref{cond2} prove useful, since they allow the higher derivatives of $\mathcal{L}(\phi)$ to be reduced to expressions involving only the function itself and the parameters $A$ and $B$. The resulting expressions are remarkably simple, depending only on the first two Dym charges $H_0$ and $H_1$, defined by
\begin{equation}
  H_0=\int_0^{2\pi}\cL(\phi)\,d\phi, \quad\quad H_1=\int_0^{2\pi}\sqrt{\cL(\phi)}\,d\phi,
\end{equation}
and exhibit a recursive structure reminiscent of the recursion relation \eqref{recursion},
\begin{subequations}
  \begin{align}
    H_2=& 2 B H_1 -2 H_0\,,\\
    H_3=&  \frac{2}{3}\left( A-3B^2   \right)H_1 +\frac{4}{3} B H_0\,,\\
    H_4=& \frac{4}{5}\left( -3 A B + 5 B^3   \right)H_1 +\frac{4}{5}\left(     A - 3 B^2   \right)H_0\,,\\
    H_5=& \frac{2}{7}\left( -3 A^2 + 30 A B^2 - 35 B^4   \right)H_1 +\frac{8}{7}\left(-  3 A B + 5 B^3      \right)H_0\,,\\
    H_6=& \frac{4}{9}\left(    15 A^2 B - 70 A B^3 + 63 B^5\right)H_1 +\frac{4}{9}\left( -3 A^2 + 30 A B^2 - 35 B^4       \right)H_0\,,\\
     H_7=&\frac{4}{11}\left(  5 A^3 - 105 A^2 B^2 + 315 A B^4 - 231 B^6 \right)H_1\nonumber\\ +&\frac{8}{11} \left( 15 A^2 B - 70 A B^3 + 63 B^5    \right) H_0\,,
  \end{align}
\end{subequations}
etc. Surprisingly, when setting $A=1$, the coefficients in parentheses correspond to the Legendre polynomials evaluated on $B$.

To finalize this section, we show that the periodic solutions indeed fulfill the conditions specified at the end of the previous section, which are necessary to describe black hole spacetimes.  

First, let us examine the condition $\LL(\phi) > 0$ required for a nontrivial angular holonomy. This is trivially satisfied within the family of periodic solutions identified here, as the function $ \cL(\phi) $ oscillates between two positive values, without ever reaching zero. 

Regarding the temporal holonomy, the inverse temperature $\beta$ can be evaluated from Eq. \eqref{beta}. As we are addressing the first nontrivial case of the Dym boundary conditions with $N=1$, the function $\mu$ corresponds to
\begin{equation}
  \mu=1+\frac{12 \pi}{c}\frac{1}{ \sqrt{\cL(\phi)}}\,.
\end{equation}
The inverse temperature can then be reduced to a constant by virtue of the conditions \eqref{cond1} and \eqref{cond2}, yielding
\begin{equation}\label{betaAB}
  \beta=\sqrt{\frac{c}{12\pi}}\frac{ \ell \pi} {\sqrt{A+\frac{12 \pi}{c}(2B+\frac{12\pi}{c})}}\,,
\end{equation}
Given the constraints $A > 0$ and $B > \sqrt{A}$, $\beta$ is guaranteed to be real and positive. Therefore, the temporal holonomy is indeed trivial within the periodic solution space.

\section{Metric Formalism}\label{Metric}
Since we are dealing with two copies of the construction described in the previous section, the full parameter space of the black hole solution comprises three real numbers and two integers.

On the one hand, we have the two positive real numbers $A^\pm$ and the two integers $m^\pm$. As explained above, for a fixed value of $A^\pm$ only finitely many values of $m^\pm$ are allowed (see Fig. \ref{AB}). These two parameters uniquely determine the corresponding values of $B^\pm$ through Eq. \eqref{period}.

Regarding the phases $\phi_0^\pm$ introduced in \eqref{final_integration}, only one of them can be set to zero by a rotation of the reference frame. Hence, the physically relevant parameter is the phase difference $\Delta \phi\equiv \phi_0^+-\phi_0^-$, which is constrained to lie within $0 \leq \Delta \phi < 2\pi$.

Once these five parameters are fixed, physical quantities of the black hole configuration can be determined \cite{Banados:1994tn}. For instance, the black hole mass $M$ is given by
\begin{equation}
  M=\frac{1}{\ell}\left(H_0^++H_0^-+\frac{24\pi}{c}\left(H_1^++H_1^- \right)     \right)\,,
\end{equation}
which is obtained as the boundary term necessary to obtain a functionally differentiable Hamiltonian~\cite{Regge:1974zd}. 

Although $\partial_\phi$ is not a Killing vector for the solution, the angular momentum of the black hole can still be obtained from the boundary term of the generator associated with the corresponding asymptotic symmetry~\cite{Lara:2024cie}. Its value is
\begin{equation}\label{J}
  J=H_0^+-H_0^-\,.
\end{equation}

The inverse temperatures of both sectors $\beta^\pm$ can be computed from \eqref{betaAB}, while the total black hole inverse temperature is given by $\beta = (\beta^+ + \beta^-)/2$.

The metric field is constructed from the connections $\mathcal A^\pm$ \eqref{A} by the relation
\begin{equation}
  g_{\mu\nu}=\frac{\ell^2}{2}\left< (\mathcal A^+_\mu-\mathcal A^-_\mu),(\mathcal A^+_\nu-\mathcal A^-_\nu) \right>\,,
\end{equation}
where $\left<\, ,\,\right>$ denotes the invariant bilinear form of $\mathfrak{sl}(2,\mathbb{R})$. The radial dependence is restored by choosing the group element $b_\pm=\exp\left[\pm \log \left(\frac{\rho}{\ell}    \right) L_0 \right] $ in \eqref{A}. The nonvanishing components of the metric then read
\begin{subequations}\label{gmunu}
  \begin{align}	
    g_{tt}=&-\frac{\rho^2}{\ell^2}\left(1+\frac{12\pi}{c\sqrt{\LL^+}}\right)\left(1+\frac{12\pi}{c\sqrt{\LL^-}}\right)+\frac{6\pi}{c}\left[2\left(A^++A^-\right)+\frac{48\pi}{c}\left(B^++B^-\right)\right.\nonumber\\
           &\left.+\frac{576\pi^2}{c^2}-\frac{3\pi}{c}\frac{{\LL^+}'{\LL^-}'}{{\LL^+}^{3/2}{\LL^-}^{3/2}}\right]-\frac{144\pi^2\ell^2}{c^2\rho^2}\left(A^++\frac{12\pi}{c}\sqrt{\LL^+}\right)\left(A^-+\frac{12\pi}{c}\sqrt{\LL^-}\right)\label{gtt}\\
    g_{t\rho}=&\frac{3\pi\ell}{c\rho}\left(\frac{{\LL^+}'}{{\LL^+}^{3/2}}-\frac{{\LL^-}'}{{\LL^-}^{3/2}}\right)\label{gtrho}\\
    g_{t\phi}=&\frac{6\pi\rho^2}{c\ell}\left(\frac{1}{\sqrt{\LL^+}}-\frac{1}{\sqrt{\LL^-}}\right)+\frac{6\pi\ell}{c}\left[\left(\LL^++A^+\right)-\left(\LL^-+A^-\right)+\frac{24\pi}{c}\left(\sqrt{\LL^+}-\sqrt{\LL^-}\right)\right]\nonumber\\
           &+\frac{72\pi^2\ell^3}{c^2\rho^2}\left[\left(A^++\frac{12\pi}{c}\sqrt{\LL^+}\right)\LL^--\left(A^-+\frac{12\pi}{c}\sqrt{\LL^-}\right)\LL^+\right]\label{gtphi}\\
    g_{\rho\rho}=&\frac{\ell^2}{\rho^2}\\
    g_{\phi\phi}=&\rho^2 +\frac{12\pi\ell^2}{c}\left(\LL^++\LL^-\right)+\frac{144\pi^2\ell^4}{c^2\rho^2}\LL^+\LL^-
  \end{align}
\end{subequations}
In the above expressions, primes denote derivatives with respect to $\phi$. The relations \eqref{cond1} and \eqref{cond2} are helpful to express the metric in the desired form, which involves the functions $\cL^\pm(\phi)$ and the parameters \(A^\pm\) and \(B^\pm\). We intentionally leave the first derivatives appearing in \eqref{gtt} and \eqref{gtrho} unchanged, to avoid introducing $\pm$ signs.

Note that the above metric does not satisfy the standard Brown-Henneaux boundary conditions \cite{Brown:1986nw}. Instead, it lies within a broader class of asymptotic behaviors \cite{Ciambelli:2019bzz,Alessio:2020ioh,Jia:2021hgy,Ciambelli:2023ott,Arenas-Henriquez:2024ypo}, which generalizes the Fefferman-Graham expansion in three dimensions \cite{Fefferman:2007rka,fefferman1985elie,Banados:2004nr,Skenderis:1999nb}.

When evaluated in the above metric, Einstein's equations are solved by virtue of Dym equation \eqref{stationary_dym} and the relations \eqref{cond1} and \eqref{cond2}.

\begin{figure}[h]
  \centering
  \begin{subfigure}[b]{0.45\linewidth}
    \begin{center}
  \begin{tikzpicture}[scale=0.11, every node/.style={scale=0.8}]
   
    \def\data{ (33.7681, 0) (25.4694, 0.800407) (15.55, 0.978324) (10.0219, 0.947349) (7.07488, 0.893766) (5.43117, 0.860212) (4.48157, 0.854905) (3.93752, 0.880141) (3.66006, 0.939743) (3.58658, 1.042) (3.70135, 1.20264) (4.02782, 1.4501) (4.63638, 1.83567) (5.66891, 2.45316) (7.38294, 3.47415) (10.2014, 5.19788) (14.7024, 8.08274) (21.2782, 12.5839) (26.128, 16.5814) (19.292, 13.1108) (11.2834, 8.19784) (7.11641, 5.52005) (5.00397, 4.13964) (3.85316, 3.39702) (3.19472, 3.00004) (2.81992, 2.81992) (2.63195, 2.80274) (2.58861, 2.9362) (2.68021, 3.23982) (2.92262, 3.76782) (3.35619, 4.6194) (4.0398, 5.94438) (5.01485, 7.90214) (6.23343, 10.5402) (7.62247, 13.8652) (9.36543, 18.3807) (10.8313, 23.0178) (8.27158, 19.1145) (4.72495, 11.9339) (2.85605, 7.93298) (1.91973, 5.90834) (1.40541, 4.83745) (1.09768, 4.27517) (0.901025, 4.03096) (0.768396, 4.02807) (0.673365, 4.25146) (0.597499, 4.72969) (0.522291, 5.52526) (0.42168, 6.70241) (0.258404, 8.22256) (0, 9.82593) (-0.354976, 11.2955) (-0.820391, 13.0397) (-1.52056, 16.0858) (-2.58214, 20.4397) (-2.97785, 18.8014) (-2.31969, 12.1602) (-1.81037, 8.09913) (-1.5519, 6.04425) (-1.44525, 4.97459) (-1.43982, 4.43132) (-1.51964, 4.22096) (-1.68978, 4.26789) (-1.97129, 4.55538) (-2.39532, 5.09031) (-2.97869, 5.846) (-3.66107, 6.65946) (-4.27703, 7.23206) (-4.74221, 7.47253) (-5.23649, 7.70527) (-6.11436, 8.41569) (-7.88728, 10.1682) (-10.9338, 13.2167) (-11.8913, 13.4881) (-8.57814, 9.13479) (-6.06755, 6.06755) (-4.79395, 4.50181) (-4.19062, 3.69453) (-3.97652, 3.28966) (-4.04154, 3.13494) (-4.35318, 3.16277) (-4.89963, 3.32979) (-5.61159, 3.56123) (-6.2655, 3.70541) (-6.5832, 3.61914) (-6.5885, 3.35701) (-6.60007, 3.10576) (-6.93906, 3.0028) (-7.92407, 3.13736) (-10.0922, 3.63342) (-14.1303, 4.59123) (-16.819, 4.88636) (-12.8632, 3.30271) (-9.08917, 2.03167) (-7.14601, 1.36317) (-6.2488, 0.989713) (-5.95364, 0.75212) (-6.06646, 0.573449) (-6.46411, 0.406688) (-6.9472, 0.218324) (-7.17875, 0) (-6.95256, -0.218493) (-6.47439, -0.407334) (-6.08174, -0.574894) (-5.97536, -0.754863) (-6.28072, -0.994768) (-7.19661, -1.37283) (-9.17663, -2.05122) (-13.0034, -3.33871) (-16.8436, -4.89351) (-13.9849, -4.54398) (-9.99151, -3.59717) (-7.86507, -3.114) (-6.90167, -2.98662) (-6.57458, -3.09376) (-6.57082, -3.348) (-6.57193, -3.61295) (-6.26013, -3.70223) (-5.61129, -3.56103) (-4.90328, -3.33227) (-4.36013, -3.16782) (-4.05192, -3.14299) (-3.99142, -3.30199) (-4.21268, -3.71397) (-4.82897, -4.53471) (-6.12826, -6.12826) (-8.68184, -9.24522) (-11.9464, -13.5506) (-10.822, -13.0815) (-7.79783, -10.0529) (-6.06176, -8.3433) (-5.20385, -7.65723) (-4.72101, -7.43912) (-4.2637, -7.20951) (-3.65371, -6.64607) (-2.97557, -5.83988) (-2.39497, -5.08957) (-1.97278, -4.55883) (-1.69269, -4.27525) (-1.52394, -4.2329) (-1.44582, -4.44978) (-1.45374, -5.00382) (-1.56458, -6.09362) (-1.83073, -8.19022) (-2.35265, -12.333) (-3.00339, -18.9627) (-2.55575, -20.2308) (-1.49953, -15.8634) (-0.811578, -12.8997) (-0.352207, -11.2074) (0, -9.77095) (0.257392, -8.19033) (0.420597, -6.6852) (0.521562, -5.51755) (0.597319, -4.72827) (0.673906, -4.25487) (0.769924, -4.03608) (0.904034, -4.04441) (1.10313, -4.2964) (1.4153, -4.87148) (1.93854, -5.96622) (2.89472, -8.04039) (4.80833, -12.1445) (8.38697, -19.3811) (10.7264, -22.7947) (9.1983, -18.0527) (7.51319, -13.6664) (6.16831, -10.43) (4.97717, -7.84277) (4.01837, -5.91285) (3.34428, -4.60301) (2.91663, -3.7601) (2.67842, -3.23765) (2.59041, -2.93824) (2.63758, -2.80874) (2.83052, -2.83052) (3.21289, -3.0171) (3.88442, -3.42458) (5.06062, -4.18651) (7.22805, -5.60665) (11.5192, -8.36916) (19.6728, -13.3697) (25.9494, -16.468) (20.814, -12.3093) (14.4318, -7.93395) (10.0637, -5.1277) (7.31107, -3.44033) (5.62971, -2.43619) (4.61461, -1.82706) (4.01646, -1.44602) (3.69723, -1.2013) (3.58857, -1.04257) (3.66846, -0.941901) (3.9542, -0.883869) (4.5108, -0.860481) (5.48205, -0.868272) (7.16765, -0.905484) (10.2044, -0.964596) (15.9329, -1.00242) (26.1383, -0.82143) (33.7206, 0)}

    \draw [] plot [smooth] coordinates {\data};

  \end{tikzpicture}
\end{center}
\caption{$(A^+,A^-)=(11.2,11.2)$, $(m^+,m^-)=(12,11)$, implying $(B^+,B^-)=(3.53,4.40)$, and $\Delta\phi=0$. Invariant under dihedral group $D_1$.}
\label{rhozero1}
\end{subfigure}\hfill
\begin{subfigure}[b]{0.45\linewidth}
  \begin{center}
  \begin{tikzpicture}[scale=0.41, every node/.style={scale=0.8}]

    \def\data{(2.48598, 0) (2.39755, 0.0753459) (2.28374, 0.143681) (2.15462, 0.203672) (2.02142, 0.255365) (1.89355, 0.299908) (1.77735, 0.339048) (1.67636, 0.374711) (1.59203, 0.408765) (1.52464, 0.442949) (1.47393, 0.478909) (1.43961, 0.51829) (1.42161, 0.562857) (1.4204, 0.614664) (1.43714, 0.676265) (1.47396, 0.751021) (1.53441, 0.843549) (1.624, 0.960428) (1.75119, 1.11134) (1.9289, 1.31088) (2.1766, 1.58139) (2.52231, 1.95651) (2.99946, 2.48137) (3.61737, 3.18914) (4.25449, 3.99524) (4.53615, 4.53615) (4.18742, 4.45914) (3.50386, 3.97435) (2.85861, 3.45547) (2.36419, 3.0479) (2.00498, 2.75962) (1.74399, 2.5662) (1.5509, 2.44383) (1.40426, 2.37448) (1.28898, 2.34464) (1.1939, 2.34316) (1.1102, 2.3593) (1.03058, 2.38153) (0.949155, 2.39729) (0.862047, 2.39443) (0.768106, 2.36399) (0.669121, 2.30313) (0.56897, 2.21599) (0.47203, 2.11174) (0.381723, 2.00106) (0.299862, 1.89325) (0.226749, 1.7949) (0.161636, 1.70993) (0.1032, 1.64032) (0.0498687, 1.58685) (0, 1.5498) (-0.048063, 1.52939) (-0.0960155, 1.52612) (-0.145673, 1.54106) (-0.199108, 1.5761) (-0.258854, 1.63434) (-0.328214, 1.72056) (-0.411749, 1.84206) (-0.516056, 2.00991) (-0.651048, 2.24092) (-0.831982, 2.56058) (-1.08222, 3.00597) (-1.43439, 3.62285) (-1.91703, 4.42999) (-2.48666, 5.28443) (-2.91256, 5.71622) (-2.94528, 5.35745) (-2.69487, 4.55678) (-2.40103, 3.78342) (-2.16661, 3.18807) (-2.0038, 2.75799) (-1.90078, 2.45047) (-1.84428, 2.22935) (-1.82379, 2.06868) (-1.83116, 1.94999) (-1.85923, 1.85923) (-1.90052, 1.7847) (-1.9463, 1.7159) (-1.98647, 1.64335) (-2.01066, 1.55963) (-2.01073, 1.46088) (-1.98343, 1.34794) (-1.93152, 1.22578) (-1.86236, 1.1014) (-1.78507, 0.981348) (-1.70791, 0.870224) (-1.63706, 0.770343) (-1.57648, 0.682205) (-1.52845, 0.605157) (-1.49421, 0.53795) (-1.47454, 0.479106) (-1.47014, 0.427116) (-1.48205, 0.380526) (-1.51184, 0.337936) (-1.56197, 0.297962) (-1.63619, 0.259147) (-1.74011, 0.219827) (-1.88212, 0.177913) (-2.07488, 0.130541) (-2.33757, 0.0734612) (-2.69932, 0) (-3.20285, -0.100654) (-3.90177, -0.245478) (-4.82145, -0.455761) (-5.80751, -0.733659) (-6.33666, -1.00363) (-5.99211, -1.14306) (-5.15009, -1.15118) (-4.32633, -1.11081) (-3.69098, -1.07233) (-3.23428, -1.05088) (-2.91188, -1.04834) (-2.68548, -1.06326) (-2.52733, -1.09367) (-2.41745, -1.13757) (-2.34036, -1.19247) (-2.28267, -1.25491) (-2.23171, -1.31983) (-2.17539, -1.38055) (-2.10346, -1.42951) (-2.00974, -1.46016) (-1.89404, -1.46917) (-1.76197, -1.45763) (-1.62245, -1.43038) (-1.48453, -1.39407) (-1.35514, -1.35514) (-1.23837, -1.31873) (-1.13594, -1.28847) (-1.04791, -1.26671) (-0.973505, -1.25503) (-0.911541, -1.25463) (-0.860819, -1.26666) (-0.820267, -1.29253) (-0.78905, -1.33421) (-0.766614, -1.39447) (-0.752737, -1.47733) (-0.747578, -1.58868) (-0.751755, -1.7372) (-0.766461, -1.93586) (-0.793604, -2.20432) (-0.835854, -2.57249) (-0.896143, -3.08455) (-0.974831, -3.79672) (-1.05923, -4.73872) (-1.09886, -5.76041) (-1.00364, -6.3367) (-0.76321, -6.04143) (-0.495381, -5.24058) (-0.279759, -4.44665) (-0.12046, -3.83311) (0, -3.39424) (0.0970557, -3.08836) (0.181115, -2.87875) (0.258869, -2.73855) (0.334548, -2.64822) (0.410586, -2.59234) (0.487801, -2.55714) (0.565309, -2.52905) (0.640481, -2.49451) (0.709306, -2.44145) (0.76746, -2.362) (0.811839, -2.25497) (0.841752, -2.12602) (0.859055, -1.98516) (0.867264, -1.84303) (0.870374, -1.7082) (0.872008, -1.58618) (0.87508, -1.47968) (0.881821, -1.38953) (0.893979, -1.31545) (0.91306, -1.25672) (0.940556, -1.21256) (0.978169, -1.1824) (1.02804, -1.16608) (1.09303, -1.16396) (1.17711, -1.17711) (1.28592, -1.20755) (1.42767, -1.25866) (1.61459, -1.33571) (1.86517, -1.44677) (2.20761, -1.60393) (2.68373, -1.82386) (3.34699, -2.12406) (4.22558, -2.499) (5.17671, -2.84592) (5.71369, -2.91127) (5.46649, -2.57234) (4.77811, -2.06767) (4.09726, -1.62222) (3.57292, -1.28633) (3.2008, -1.04) (2.94578, -0.855828) (2.77665, -0.712922) (2.67046, -0.596919) (2.6103, -0.497941) (2.58249, -0.409026) (2.57436, -0.325217) (2.57285, -0.243206) (2.56431, -0.161333) (2.53609, -0.0796999) (2.4794, 0) }

    \draw [] plot [smooth] coordinates {\data};

  \end{tikzpicture}
\end{center}
\caption{$(A^+,A^-)=(3.50,4.00)$, $(m^+,m^-)=(5,5)$, implying $(B^+,B^-)=(1.99,2.74)$, and $\Delta\phi=\pi/4$. Invariant under cyclic group $C_5$.}
\label{rhozero2}
\end{subfigure}
\begin{subfigure}[b]{0.45\linewidth}
  \begin{center}
  \begin{tikzpicture}[scale=2.2, every node/.style={scale=0.8}]

    \def\data{ (1.19535, 0) (1.22372, 0.038457) (1.23947, 0.0779811) (1.24357, 0.117552) (1.23795, 0.15639) (1.22513, 0.194041) (1.2077, 0.23038) (1.18805, 0.26556) (1.16819, 0.299941) (1.14974, 0.334031) (1.13392, 0.368433) (1.12166, 0.403821) (1.11366, 0.440928) (1.11047, 0.480541) (1.11252, 0.523512) (1.12015, 0.570745) (1.13356, 0.62318) (1.15273, 0.681722) (1.17721, 0.747081) (1.20581, 0.819467) (1.23604, 0.898038) (1.2635, 0.980068) (1.28127, 1.05996) (1.2803, 1.12873) (1.25139, 1.17513) (1.18924, 1.18924) (1.09616, 1.1673) (0.982016, 1.11388) (0.8599, 1.03944) (0.741118, 0.955443) (0.632691, 0.870824) (0.537575, 0.791018) (0.456024, 0.718579) (0.386905, 0.65422) (0.328565, 0.597658) (0.279303, 0.548164) (0.237575, 0.504872) (0.202058, 0.466928) (0.171657, 0.433556) (0.145479, 0.404083) (0.122797, 0.37793) (0.103025, 0.354615) (0.085687, 0.333729) (0.0703956, 0.314932) (0.0568347, 0.297938) (0.0447448, 0.282508) (0.0339116, 0.268438) (0.0241574, 0.255558) (0.0153338, 0.243723) (0.00731628, 0.232808) (0, 0.222706) (-0.00670408, 0.213327) (-0.0128718, 0.204591) (-0.018568, 0.196429) (-0.0238487, 0.188782) (-0.028762, 0.181596) (-0.0333499, 0.174826) (-0.037649, 0.168432) (-0.0416911, 0.162376) (-0.0455045, 0.156627) (-0.049114, 0.151157) (-0.0525416, 0.14594) (-0.055807, 0.140952) (-0.0589277, 0.136174) (-0.0619196, 0.131586) (-0.0647968, 0.127171) (-0.0675723, 0.122914) (-0.0702579, 0.1188) (-0.0728643, 0.114816) (-0.0754016, 0.11095) (-0.0778789, 0.107191) (-0.0803048, 0.103528) (-0.0826874, 0.0999519) (-0.0850343, 0.0964525) (-0.0873529, 0.0930214) (-0.08965, 0.08965) (-0.0919325, 0.0863304) (-0.0942069, 0.0830546) (-0.0964798, 0.079815) (-0.0987575, 0.0766041) (-0.101046, 0.0734146) (-0.103353, 0.0702388) (-0.105685, 0.0670695) (-0.108047, 0.0638989) (-0.110448, 0.0607194) (-0.112895, 0.0575228) (-0.115395, 0.0543008) (-0.117957, 0.0510447) (-0.12059, 0.047745) (-0.123303, 0.0443918) (-0.126106, 0.0409744) (-0.129011, 0.0374812) (-0.13203, 0.0338995) (-0.135176, 0.0302154) (-0.138464, 0.0264135) (-0.141911, 0.0224765) (-0.145536, 0.0183854) (-0.149358, 0.0141185) (-0.153402, 0.00965123) (-0.157694, 0.00495573) (-0.162264, 0) (-0.167147, -0.00525281) (-0.172382, -0.0108454) (-0.178016, -0.0168275) (-0.1841, -0.0232573) (-0.190697, -0.0302034) (-0.197877, -0.0377471) (-0.205726, -0.0459852) (-0.21434, -0.0550333) (-0.223837, -0.0650306) (-0.234351, -0.0761454) (-0.246044, -0.0885812) (-0.259101, -0.102585) (-0.273739, -0.118457) (-0.290201, -0.136558) (-0.308756, -0.157319) (-0.329672, -0.181239) (-0.353182, -0.208871) (-0.379399, -0.240774) (-0.40816, -0.277385) (-0.438781, -0.318793) (-0.469692, -0.36433) (-0.498061, -0.412032) (-0.519694, -0.458172) (-0.529725, -0.497445) (-0.524357, -0.524357) (-0.502888, -0.535521) (-0.468334, -0.531221) (-0.425969, -0.514908) (-0.381057, -0.491255) (-0.337412, -0.464407) (-0.297152, -0.437246) (-0.261104, -0.411434) (-0.229321, -0.387761) (-0.201474, -0.36648) (-0.177089, -0.347557) (-0.155671, -0.330819) (-0.136764, -0.316042) (-0.119965, -0.302997) (-0.104934, -0.291466) (-0.0913848, -0.281253) (-0.0790783, -0.272189) (-0.067816, -0.264126) (-0.0574323, -0.256937) (-0.0477884, -0.250515) (-0.0387676, -0.244769) (-0.0302709, -0.239619) (-0.0222139, -0.234999) (-0.0145239, -0.230851) (-0.00713774, -0.227126) (0, -0.223782) (0.00693835, -0.220782) (0.0137213, -0.218094) (0.0203887, -0.21569) (0.0269773, -0.213548) (0.0335212, -0.211644) (0.0400526, -0.209963) (0.0466023, -0.208487) (0.0532005, -0.207202) (0.0598766, -0.206097) (0.0666602, -0.205159) (0.0735814, -0.20438) (0.0806708, -0.203751) (0.0879604, -0.203265) (0.0954838, -0.202913) (0.103277, -0.202692) (0.111377, -0.202593) (0.119825, -0.202614) (0.128667, -0.202747) (0.137951, -0.202989) (0.147731, -0.203334) (0.158066, -0.203777) (0.169023, -0.204313) (0.180674, -0.204935) (0.193104, -0.205635) (0.206405, -0.206405) (0.220681, -0.207233) (0.23605, -0.208106) (0.252646, -0.209007) (0.270619, -0.209913) (0.29014, -0.210799) (0.311403, -0.211629) (0.334622, -0.212358) (0.360043, -0.212929) (0.387935, -0.213269) (0.418593, -0.213284) (0.452338, -0.212854) (0.489502, -0.211826) (0.530419, -0.210008) (0.575396, -0.207155) (0.624669, -0.202967) (0.678348, -0.197078) (0.736325, -0.189056) (0.798163, -0.178411) (0.862974, -0.164621) (0.929297, -0.147186) (0.995039, -0.125703) (1.05754, -0.0999671) (1.11382, -0.0700753) (1.16098, -0.0364852) (1.19678, 0) }

    \draw [] plot [smooth] coordinates {\data};

  \end{tikzpicture}
\end{center}
\caption{$(A^+,A^-)=(0.5,1.2)$, $(m^+,m^-)=(1,2)$, implying $(B^+,B^-)=(1.09,1.55)$, and $\Delta\phi=\pi/4$. No discrete symmetry group.}
\label{rhozero3}
\end{subfigure}\hfill
\begin{subfigure}[b]{0.45\linewidth}
  \begin{center}
  \begin{tikzpicture}[scale=0.5, every node/.style={scale=0.8}]

    \def\data{ (1.6923, 0) (1.6839, 0.0529186) (1.65982, 0.104427) (1.62319, 0.153437) (1.57825, 0.199379) (1.52945, 0.242241) (1.48076, 0.282471) (1.43537, 0.320844) (1.39562, 0.358334) (1.36318, 0.39604) (1.33927, 0.435156) (1.32489, 0.47699) (1.32101, 0.523023) (1.32876, 0.575006) (1.34966, 0.635102) (1.3858, 0.706102) (1.44016, 0.791733) (1.51694, 0.897114) (1.62212, 1.02943) (1.76405, 1.19885) (1.9538, 1.41952) (2.20403, 1.70962) (2.52195, 2.08634) (2.88562, 2.54401) (3.19534, 3.00062) (3.26539, 3.26539) (3.00066, 3.19538) (2.54407, 2.88569) (2.08642, 2.52205) (1.7097, 2.20414) (1.41961, 1.95392) (1.19894, 1.76418) (1.02952, 1.62226) (0.897204, 1.51709) (0.791822, 1.44032) (0.706191, 1.38598) (0.635191, 1.34985) (0.575094, 1.32896) (0.52311, 1.32122) (0.477074, 1.32513) (0.435237, 1.33952) (0.396118, 1.36345) (0.358406, 1.3959) (0.320909, 1.43566) (0.282527, 1.48106) (0.242286, 1.52973) (0.199412, 1.57851) (0.153458, 1.62341) (0.104437, 1.65998) (0.0529212, 1.68398) (0, 1.6923) (-0.0529158, 1.68381) (-0.104417, 1.65966) (-0.153415, 1.62297) (-0.199346, 1.57798) (-0.242195, 1.52916) (-0.282414, 1.48046) (-0.320777, 1.43507) (-0.35826, 1.39533) (-0.39596, 1.3629) (-0.435071, 1.33901) (-0.476901, 1.32464) (-0.522932, 1.32077) (-0.574912, 1.32854) (-0.635005, 1.34945) (-0.706003, 1.38561) (-0.79163, 1.43997) (-0.897008, 1.51676) (-1.02932, 1.62195) (-1.19873, 1.76388) (-1.41939, 1.95363) (-1.70949, 2.20386) (-2.0862, 2.52178) (-2.54388, 2.88546) (-3.00053, 3.19524) (-3.26539, 3.26539) (-3.19548, 3.00075) (-2.88584, 2.54421) (-2.52222, 2.08656) (-2.20431, 1.70984) (-1.95409, 1.41973) (-1.76436, 1.19906) (-1.62244, 1.02963) (-1.51727, 0.897311) (-1.44051, 0.791925) (-1.38617, 0.706291) (-1.35006, 0.635288) (-1.32918, 0.575189) (-1.32146, 0.523203) (-1.32538, 0.477165) (-1.33979, 0.435324) (-1.36373, 0.3962) (-1.3962, 0.358484) (-1.43598, 0.320979) (-1.48138, 0.282588) (-1.53005, 0.242336) (-1.57881, 0.19945) (-1.62368, 0.153483) (-1.6602, 0.104451) (-1.68414, 0.0529262) (-1.69239, 0) (-1.68382, -0.0529163) (-1.65961, -0.104414) (-1.62288, -0.153407) (-1.57787, -0.199331) (-1.52904, -0.242176) (-1.48036, -0.282393) (-1.43499, -0.320758) (-1.39528, -0.358246) (-1.36289, -0.395956) (-1.33904, -0.435081) (-1.32473, -0.476931) (-1.32092, -0.522987) (-1.32875, -0.575001) (-1.34974, -0.635138) (-1.38598, -0.706192) (-1.44045, -0.791896) (-1.51738, -0.897376) (-1.62275, -1.02983) (-1.76491, -1.19943) (-1.95496, -1.42036) (-2.20556, -1.71081) (-2.52387, -2.08793) (-2.88768, -2.54583) (-3.19683, -3.00202) (-3.26539, -3.26539) (-2.99926, -3.19389) (-2.54226, -2.88362) (-2.08484, -2.52014) (-1.70852, -2.20261) (-1.41876, -1.95276) (-1.19836, -1.76333) (-1.02913, -1.62164) (-0.896944, -1.51665) (-0.79166, -1.44002) (-0.706102, -1.3858) (-0.635155, -1.34977) (-0.575098, -1.32897) (-0.523145, -1.32131) (-0.477132, -1.32528) (-0.43531, -1.33975) (-0.396198, -1.36372) (-0.35849, -1.39623) (-0.320989, -1.43602) (-0.282598, -1.48143) (-0.242343, -1.5301) (-0.199452, -1.57883) (-0.15348, -1.62364) (-0.104444, -1.66009) (-0.0529195, -1.68392) (0, -1.69206) (0.0529023, -1.68338) (0.104379, -1.65905) (0.153345, -1.62222) (0.199238, -1.57713) (0.242048, -1.52823) (0.282228, -1.47949) (0.320553, -1.43407) (0.357996, -1.3943) (0.395656, -1.36186) (0.434724, -1.33794) (0.476506, -1.32355) (0.522478, -1.31963) (0.574388, -1.32733) (0.634394, -1.34816) (0.705278, -1.38419) (0.790757, -1.43838) (0.895935, -1.51494) (1.02797, -1.61983) (1.19701, -1.76135) (1.41716, -1.95055) (1.70657, -2.2001) (2.08253, -2.51735) (2.53981, -2.88085) (2.99747, -3.19198) (3.26538, -3.26538) (3.19872, -3.0038) (2.89046, -2.54828) (2.52667, -2.09024) (2.20808, -1.71276) (1.95718, -1.42198) (1.7669, -1.20078) (1.62457, -1.03098) (1.51909, -0.898386) (1.4421, -0.792799) (1.3876, -0.707015) (1.35135, -0.635897) (1.33038, -0.575707) (1.32258, -0.523648) (1.32645, -0.47755) (1.34081, -0.435656) (1.36471, -0.396485) (1.39713, -0.358723) (1.43685, -0.321175) (1.48218, -0.282741) (1.53074, -0.242446) (1.57936, -0.199519) (1.62403, -0.153516) (1.6603, -0.104458) (1.68395, -0.0529202) (1.69188, 0) }

    \draw [] plot [smooth] coordinates {\data};

  \end{tikzpicture}
\end{center}
\caption{$(A^+,A^-)=(2.60,3.00)$, $(m^+,m^-)=(4,4)$, implying $(B^+,B^-)=(1.72,2.42)$, and $\Delta\phi=\pi/4$. Invariant under dihedral group $D_4$.}
\label{rhozero4}
\end{subfigure}
\caption{Examples of surfaces at fixed $\rho = \rho_0 \ll \ell$, illustrating the nonaxisymmetric features of the spacetime. The plots are not to scale with respect to each other.}
\label{rhozerosurfaces}
\end{figure}
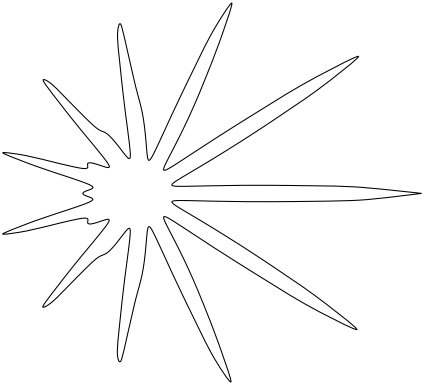
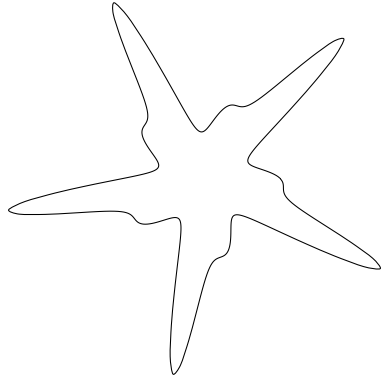
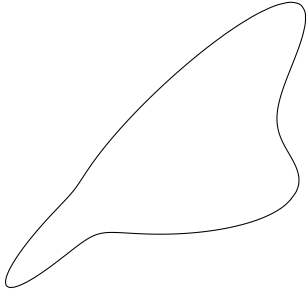
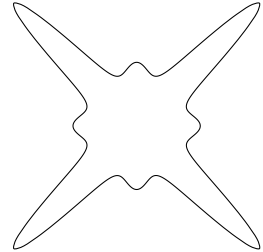

To visualize the nonaxisymmetric features of the spacetime, a useful approach is to examine surfaces at fixed time and radial coordinate, $\rho = \rho_0$. For simplicity, we focus on surfaces close to the origin, i.e., with $\rho_0 \ll \ell$. The induced geometry on such a surface is approximately given by
\begin{equation}
  dl \approx \frac{12\pi\ell^2}{c\rho_0} \,\sqrt{ \LL^+(\phi) \LL^-(\phi)} \, d\phi \, .
\end{equation}
These surfaces correspond to closed curves which wobble around the origin. Several examples for different parameter choices are illustrated in Fig. \ref{rhozerosurfaces}. By appropriately tuning the parameters, the resulting curves can display discrete rotational symmetries, reflective symmetries, or a combination of both.

The absence of axisymmetry in the present solutions renders the identification of the black hole horizon significantly more delicate. Stationary nonaxisymmetric black holes are comparatively rare in the literature, mainly due to four-dimensional uniqueness theorems~\cite{PhysRev.164.1776,israel1968event,mazur1982proof,mazur1984black, PhysRevLett.26.331,hawking1972black,hawking1973event,bardeen1973four, PhysRevLett.34.905,hawking2023large}. Explicit examples are typically found in higher or lower dimensions, or with different asymptotic behavior \cite{PhysRevLett.88.101101, Cardenas:2025qqi, Barnich:2015dvt, Alkac:2016xlr, Dias:2011at, Sheikh-Jabbari:2016npa, Stotyn:2011ns, Afshar:2016wfy, Ojeda:2019xih, Banados:1998gg}.

In our case, the spacetime admits the exact Killing vector field $\partial_t$, which allows us to identify the hypersurface $\rho=\rho_K(\phi)$ where its norm vanishes. Likewise, the hypersurface where the lapse function $N=1/\sqrt{-g^{tt}}$ vanishes can be written as $\rho=\rho_N(\phi)$. The explicit forms of the functions $\rho_K(\phi)$ and $\rho_N(\phi)$, though straightforward to derive from \eqref{gmunu}, are not particularly illuminating and are therefore omitted. In general, these two hypersurfaces do not coincide, and neither provides by itself an immediate identification of the event horizon.

These difficulties are alleviated when we consider a simpler sector of the solution space. Let $A^+=A^-\equiv A$, $B^+=B^-\equiv B$ and $\LL^+=\LL^-\equiv \LL$. The metric then reduces to the diagonal form
\begin{equation}\label{gmunudiagonal}
ds^2=
-\left(\frac{12\pi}{c}\right)^2
\left[
\frac{\rho}{\ell}
\left(
\frac{c}{12\pi}+\frac{1}{\sqrt{\LL}}
\right)
-\frac{\ell}{\rho}
\left(
A+\frac{12\pi}{c}\sqrt{\LL}
\right)
\right]^2 dt^2
+\frac{\ell^2}{\rho^2}d\rho^2
+\left(
\rho+\frac{12\pi\ell^2}{c\rho}\LL
\right)^2d\phi^2.
\end{equation}

In this sector, the angular momentum \eqref{J} vanishes, while the remaining Dym charges remain nonzero. Despite this simplification, the spacetime remains nonaxisymmetric, since $\LL$ retains a nontrivial dependence on the angular coordinate $\phi$, and consequently $\partial_\phi$ is not an exact Killing vector.

For the diagonal metric, the two hypersurfaces introduced above coincide and yield the same hypersurface,
\begin{equation}\label{horizon}
\frac{\rho^2}{\ell^2}
=
\frac{12\pi}{c}
\frac{
\left( A+\frac{12\pi}{c}\sqrt{\LL}\right)
}{
\left( 1+\frac{12\pi}{c}\frac{1}{\sqrt{\LL}}\right)
}.
\end{equation}
Thus, in this particular sector, $\rho_K(\phi)=\rho_N(\phi)\equiv\rho_H(\phi)$.

The identification of \eqref{horizon} as the black hole horizon can be tested through two independent thermodynamic consistency checks. First, its surface gravity can be obtained directly from \eqref{gmunudiagonal} by the usual relation
\begin{equation}
\kappa_H^2
=
\left.
h^{ij}\partial_iN\partial_jN
\right|_{\rho=\rho_H(\phi)},
\end{equation}
where $h_{ij}$ denotes the metric on a constant-time spatial slice. Making use of \eqref{conditions}, the right-hand side becomes independent of $\phi$ and yields
\begin{equation}
\kappa_H
=
\frac{2}{\ell}
\sqrt{
\frac{12\pi}{c}
\left[
A+\frac{12\pi}{c}
\left(
2B+\frac{12\pi}{c}
\right)
\right]
},
\end{equation}
thus related to the inverse temperature $\beta$ in \eqref{betaAB} through
\begin{equation}
\kappa_H=\frac{2\pi}{\beta},
\end{equation}
as expected for a regular nonextremal Killing horizon.

A second independent check follows from the entropy. Since the horizon is described by $\rho=\rho_H(\phi)$, its length corresponds to
\begin{equation}\label{horizonarea}
\mathcal A_H
=
\int_0^{2\pi}d\phi\,
\sqrt{
\left.g_{\phi\phi}\right|_{\rho_H}
+
\left.g_{\rho\rho}\right|_{\rho_H}
\left(\rho_H'\right)^2
}.
\end{equation}
In three spacetime dimensions, $\mathcal A_H$ plays the role of the horizon area. As shown in Appendix \ref{euclideanDym}, the boundary term of the Euclidean Chern-Simons action evaluated at the horizon gives
\begin{equation}
S_{\rm CS}
=
\frac{\mathcal A_H}{4G}.
\end{equation}
Thus, the entropy obtained independently from the Chern-Simons formulation \eqref{entropy} coincides exactly with the Bekenstein-Hawking entropy associated with the geometric length of the same curve \eqref{horizon}.

The agreement between the Lorentzian surface gravity and the Euclidean temperature, together with the agreement between the Chern-Simons entropy and the geometric horizon length, provides two independent consistency checks supporting the identification of \eqref{horizon} as the black hole Killing horizon in the diagonal sector.

For the general nondiagonal solution, $\rho_K(\phi)$ and $\rho_N(\phi)$ no longer coincide, and the identification of the corresponding horizon remains more involved. A detailed analysis of this case, together with the global causal structure of the spacetime and the relation between the Killing and event horizons, will be addressed in subsequent work.

\section{Final Comments}

Having characterized the black hole spectrum for the simplest nontrivial element of the Dym family of boundary conditions, we now sketch a method for carrying out the periodic solution analysis for the higher-order equations in the Dym hierarchy.

The stationary equation for an arbitrary integer $N$ takes the form
\begin{equation}\label{dymN}
  \partial_\phi \cL^\pm - \partial_\phi^3 \left(\frac{\delta H^\pm_N}{\delta \cL^\pm}\right) = 0\,,
\end{equation}
where the Dym conserved charge $H_N^\pm$ is constructed via the recursive procedure detailed in Appendix \ref{cargasDym}. In analogy with the $N=1$ case described above, this equation admits a first integration, yielding
\begin{equation}\label{DymNintegrada}
  \cL^\pm - \partial_\phi^2 \left(\frac{\delta H^\pm_N}{\delta \cL^\pm}\right) = A^\pm\,,
\end{equation}
where the integration constant $A^\pm$ plays the same role as in \eqref{eqint2}.

To proceed with further integrations, we exploit the integrable structure of the hierarchy. As shown in Appendix \ref{proof}, multiplying the integrated equation by  $\partial_{\phi}\left(\frac{\delta H_M^{\pm}}{\delta \mathcal L^{\pm}}\right)$, where $H_M^{\pm}$ is \emph{any} of the Dym conserved quantities, transforms the equation into a total derivative, affording a further integration and introducing new integration constants. Once enough integrations have been performed for different values of $M$, finding relations equivalent to those in \eqref{cond1} and \eqref{cond2} becomes an algebraic goal, although very involved and not simple to achieve. This path toward finding black holes for the higher members of the Dym family of boundary conditions will be pursued elsewhere. 

It is worth noting that not all the integration constants emerging from the described procedure are independent, since the differential equation \eqref{dymN} is of order $2N+1$; two of them correspond to $A^{\pm}$ and the phase of the periodic solution, while the remaining $2N-1$ arise via the described integrations. Accordingly, the above procedure can only be applied for values of $M$ from $1$ to $2N-1$ before becoming redundant.  This observation can be simply illustrated in the $N=1$ case, where \eqref{DymNintegrada} becomes
\begin{equation}
\mathcal L^{\pm} - \partial_{\phi}^{2}\left(\frac{1}{2\sqrt{\mathcal L^{\pm}}}\right) = A^{\pm}\,.
\end{equation}
Multiplying by  $\partial_{\phi}\left(\frac{\delta H_{1}^{\pm}}{\delta \mathcal L^{\pm}}\right)$ yields \eqref{eqint2} and introduces the constant $B^\pm$. Multiplying instead by  $\partial_{\phi}\left(\frac{\delta H_{2}^{\pm}}{\delta \mathcal L^{\pm}}\right)$ and manipulating according to the steps described in the Appendix \ref{proof}, introduces a new constant $C^\pm$
\begin{equation}
\left(  1- \frac{\delta H_{2}^{\pm}}{\delta \mathcal L^{\pm}}  \right) \left( \d_\phi^2 \frac{\delta H_{1}^{\pm}}{\delta \mathcal L^{\pm}} - \cL^\pm \frac{\delta H_{2}^{\pm}}{\delta \mathcal L^{\pm}} \right) =A^\pm \frac{\delta H_{2}^{\pm}}{\delta \mathcal L^{\pm}}+C^\pm  \,.
\end{equation}
However, the conditions \eqref{cond1} and \eqref{cond2} imply that $C^\pm = {B^\pm}^2 - A^\pm$, so $C^\pm$ is not an independent constant. A similar observation can be made for higher values of $M$.

\paragraph{Acknowledgments} 

The authors express their gratitude to Marcela C\'ardenas, Francisco Correa, Hern\'an A. Gonz\'alez, Kristiansen Lara and Ricardo Troncoso for their insightful comments. This work has received support from FONDECYT Grant No. 1231810 and ANID Beca Doctorado Nacional No. 21232318. We also acknowledge Centro de Estudios Cient\'ificos (CECs) at Universidad San Sebasti\'an for their hospitality during part of this work.

\appendix

\section{Dym Conserved Charges $\HHpm_n$} \label{cargasDym}

As explained in~\cite{Magri1978}, the conserved quantities of the Dym hierarchy can be obtained by means of the recursion relation
\begin{align}
\label{recursion}
\cD^\pm \left(\fdv{\HHpm_{n+1}}{\LLpm}\right)=\cE\left(\fdv{\HHpm_n}{\LLpm}\right)\,,
\end{align}
where the Hamiltonian operators are given by $\cD^\pm=2 \cL^\pm \d_\phi +\d_\phi\cL^\pm  $ and $\cE=\d_\phi^3$, and considering as a starting point the simplest conserved quantity $H_0^\pm=\int d\phi\, \cL^\pm$. In what follows, the prime denotes derivatives with respect to $\phi$. Introducing the notation $\HHpm_n=\int d\phi\,\cH^\pm_n$, the first five charge densities are as follows:
\begin{align}
		\label{ap: eq: Hamiltonian densities}
		\begin{split}
			\cH^\pm_1&=\sqrt{\LLpm}\,,\\
			\cH^\pm_2&=-\frac{1}{2\sqrt{\LLpm}}\left(\frac{5}{8}\frac{{\LLpm'}^2}{\LLpm^2}-\frac12 \frac{\LLpm''}{\LLpm}\right)\,,\\
			\cH^\pm_3 &= -\frac{1}{2\sqrt{\LLpm}}\left(\frac{385}{256}\frac{{\LLpm'}^4}{\LLpm^5}-\frac{77}{32}\frac{{\LLpm'}^2\LLpm''}{\LLpm^4}+\frac{7}{16}\frac{{\LLpm''}^2}{\LLpm^3}+\frac{7}{12}\frac{\LLpm'\LLpm'''}{\LLpm^3}-\dfrac{1}{12}\frac{\LLpm^{(4)}}{\LLpm^2}\right)\,,\\
			\cH^\pm_4&=-\frac{1}{2\sqrt{\LLpm}}\left(\frac{85085}{4096}\frac{{\LLpm'}^6}{\LLpm^8}-\frac{51051}{1024}\frac{{\LLpm'}^4\LLpm''}{\LLpm^7}+\frac{35607}{1280}\frac{{\LLpm'}^2{\LLpm''}^2}{\LLpm^6}-\frac{671}{320}\frac{{\LLpm''}^3}{\LLpm^5}\right.\\
			&\hspace{2cm}+\frac{429}{32}\frac{{\LLpm'}^3\LLpm'''}{\LLpm^6} -\frac{1419}{160}\frac{\LLpm'\LLpm''\LLpm'''}{\LLpm^5}+\frac{69}{160}\frac{{\LLpm'''}^2}{\LLpm^4}-\frac{165}{64}\frac{{\LLpm'}^2\LLpm^{(4)}}{\LLpm^5}\\
			&\hspace{6cm}\left. +\frac{57}{80}\frac{\LLpm''\LLpm^{(4)}}{\LLpm^4}+\frac{27}{80}\frac{\LLpm'\LLpm^{(5)}}{\LLpm^4}-\frac{1}{40}\frac{\LLpm^{(6)}}{\LLpm^3}\right)\,,\\
			\cH^\pm_5&=-\frac{1}{2\sqrt{\mathcal{L}^\pm}}\left(\frac{185910725}{262144}\frac{ {{\mathcal{L}^\pm}'}^8}{ {\mathcal{L}^\pm}^{11}}-\frac{37182145 }{16384}\frac{{{\mathcal{L}^\pm}''} {{\mathcal{L}^\pm}'}^6}{ {\mathcal{L}^\pm}^{10}}+\frac{9006855}{14336}\frac{ {{\mathcal{L}^\pm}'''} {{\mathcal{L}^\pm}'}^5}{ {\mathcal{L}^\pm}^9}\right.\\
			&\hspace{1cm}+\frac{123740331 }{57344}\frac{{{\mathcal{L}^\pm}''}^2 {{\mathcal{L}^\pm}'}^4}{ {\mathcal{L}^\pm}^9}-\frac{1859715}{14336}\frac{ {{\mathcal{L}^\pm}^{(4)}} {{\mathcal{L}^\pm}'}^4}{ {\mathcal{L}^\pm}^8}+\frac{36465}{1792}\frac{ {{\mathcal{L}^\pm}^{(5)}} {{\mathcal{L}^\pm}'}^3}{ {\mathcal{L}^\pm}^7}\\
			&\hspace{0.5cm}-\frac{3201627}{3584}\frac{ {{\mathcal{L}^\pm}''} {{\mathcal{L}^\pm}'''} {{\mathcal{L}^\pm}'}^3}{ {\mathcal{L}^\pm}^8}+\frac{282711}{3584}\frac{ {{\mathcal{L}^\pm}'''}^2 {{\mathcal{L}^\pm}'}^2}{ {\mathcal{L}^\pm}^7}+\frac{226083}{1792}\frac{ {{\mathcal{L}^\pm}''} {{\mathcal{L}^\pm}^{(4)}} {{\mathcal{L}^\pm}'}^2}{ {\mathcal{L}^\pm}^7}\\
			&-\frac{2145 }{896}\frac{{{\mathcal{L}^\pm}^{(6)}} {{\mathcal{L}^\pm}'}^2}{ {\mathcal{L}^\pm}^6}-\frac{629629 }{1024}\frac{{{\mathcal{L}^\pm}''}^3 {{\mathcal{L}^\pm}'}^2}{ {\mathcal{L}^\pm}^8}+\frac{97383 }{448}\frac{{{\mathcal{L}^\pm}''}^2 {{\mathcal{L}^\pm}'''} {{\mathcal{L}^\pm}'}}{ {\mathcal{L}^\pm}^7}+\frac{11}{56}\frac{{{\mathcal{L}^\pm}^{(7)}} {{\mathcal{L}^\pm}'}}{ {\mathcal{L}^\pm}^5}\\
			&-\frac{1287}{112}\frac{ {{\mathcal{L}^\pm}''} {{\mathcal{L}^\pm}^{(5)}} {{\mathcal{L}^\pm}'}}{ {\mathcal{L}^\pm}^6}-\frac{3861}{224}\frac{ {{\mathcal{L}^\pm}'''} {{\mathcal{L}^\pm}^{(4)}} {{\mathcal{L}^\pm}'}}{ {\mathcal{L}^\pm}^6}+\frac{180323 }{7168}\frac{{{\mathcal{L}^\pm}''}^4}{{\mathcal{L}^\pm}^7}+\frac{253}{448}\frac{ {{\mathcal{L}^\pm}^{(4)}}^2}{{\mathcal{L}^\pm}^5}\\
			&\hspace{1cm}+\frac{209}{224}\frac{ {{\mathcal{L}^\pm}'''} {{\mathcal{L}^\pm}^{(5)}}}{{\mathcal{L}^\pm}^5}+\frac{121}{224}\frac{ {{\mathcal{L}^\pm}''} {{\mathcal{L}^\pm}^{(6)}}}{ {\mathcal{L}^\pm}^5}-\frac{1}{112}\frac{{{\mathcal{L}^\pm}^{(8)}}}{ {\mathcal{L}^\pm}^4}-\frac{13299}{896 }\frac{ {{\mathcal{L}^\pm}''} {{\mathcal{L}^\pm}'''}^2}{{\mathcal{L}^\pm}^6}\\
			&\hspace{8cm}\left.-\frac{10725}{896}\frac{ {{\mathcal{L}^\pm}''}^2 {{\mathcal{L}^\pm}^{(4)}}}{ {\mathcal{L}^\pm}^6}\right)\,.
		\end{split}
\end{align}%

The corresponding functional derivatives are
\begin{align}
	\label{ap: eq: Variational H}
\begin{split}
	\frac{\delta H_1^\pm}{\delta \cL^\pm} &= \frac{1}{2\sqrt{\LLpm}}\,,\\
	\frac{\delta H_2^\pm}{\delta \cL^\pm} &= \frac{1}{2\sqrt{\LLpm}}\left(\frac{5}{16}\frac{{\LLpm}'^2}{\LLpm^3}-\frac14 \frac{\LLpm''}{\LLpm^2}\right)\,,\\
	\frac{\delta H_3^\pm}{\delta \cL^\pm} &= \frac{1}{2\sqrt\LLpm}\left( \frac{1155}{512}\frac{{\LLpm'}^4}{ \LLpm^6}-\frac{231}{64}\frac{{\LLpm'}^2\LLpm''}{\LLpm^5}+\frac{21}{32}\frac{{\LLpm''}^2}{\LLpm^4}+\frac{7}{8}\frac{\LLpm'\LLpm'''}{\LLpm^4} - \frac{1}{8}\frac{\LLpm^{(4)}}{\LLpm^3}\right)\,,\\
	\begin{split}
	\frac{\delta H_4^\pm}{\delta \cL^\pm} &=\frac{1}{2\sqrt{\LLpm}}\left(  \frac{425425}{8192}\frac{{\LLpm'}^6}{\LLpm^9}-\frac{255255}{2048}\frac{{\LLpm'}^4\LLpm''}{\LLpm^8}+\frac{35607}{512}\frac{{\LLpm'}^2{\LLpm''}^2}{\LLpm^7}-\frac{671}{128}\frac{{\LLpm''}^3}{\LLpm^6}\right.\\
		&\hspace{1.5cm}+ \frac{2145}{64}\frac{{\LLpm'}^3\LLpm'''}{\LLpm^7}-\frac{1419}{64}\frac{{\LLpm'}{\LLpm''}{\LLpm'''}}{\LLpm^6}+\frac{69}{64}\frac{{\LLpm'''}^2}{\LLpm^5}-\frac{825}{128}\frac{{\LLpm'}^2\LLpm^{(4)}}{\LLpm^6}\\
		&\hspace{5cm}\left.+ \frac{57}{32}\frac{\LLpm''\LLpm^{(4)}}{\LLpm^5}+\frac{27}{32}\frac{\LLpm'\LLpm^{(5)}}{\LLpm^5}-\frac{1}{16}\frac{\LLpm^{(6)}}{\LLpm^4}\right)\,,\\
	\fdv{\HHpm_5}{\LLpm}&=\frac{1}{2 \sqrt{{\mathcal{L}^\pm}}} \left(\frac{1301375075}{524288}\frac{ {{\mathcal{L}^\pm}'}^8}{ {\mathcal{L}^\pm}^{12}}-\frac{260275015}{32768}\frac{ {{\mathcal{L}^\pm}''} {{\mathcal{L}^\pm}'}^6}{ {\mathcal{L}^\pm}^{11}}+\frac{9006855}{4096}\frac{ {{\mathcal{L}^\pm}'''} {{\mathcal{L}^\pm}'}^5}{ {\mathcal{L}^\pm}^{10}}\right.\\
	&\hspace{2cm}+\frac{123740331}{16384}\frac{ {{\mathcal{L}^\pm}''}^2 {{\mathcal{L}^\pm}'}^4}{ {\mathcal{L}^\pm}^{10}}-\frac{1859715}{4096}\frac{ {{\mathcal{L}^\pm}^{(4)}} {{\mathcal{L}^\pm}'}^4}{ {\mathcal{L}^\pm}^9}+\frac{36465}{512}\frac{ {{\mathcal{L}^\pm}^{(5)}} {{\mathcal{L}^\pm}'}^3}{ {\mathcal{L}^\pm}^8}\\
	&\hspace{1cm}-\frac{3201627}{1024}\frac{ {{\mathcal{L}^\pm}''} {{\mathcal{L}^\pm}'''} {{\mathcal{L}^\pm}'}^3}{ {\mathcal{L}^\pm}^9}+\frac{282711}{1024}\frac{ {{\mathcal{L}^\pm}'''}^2 {{\mathcal{L}^\pm}'}^2}{ {\mathcal{L}^\pm}^8}+\frac{226083}{512}\frac{ {{\mathcal{L}^\pm}''} {{\mathcal{L}^\pm}^{(4)}} {{\mathcal{L}^\pm}'}^2}{ {\mathcal{L}^\pm}^8}\\
	&\hspace{2cm}-\frac{2145}{256}\frac{ {{\mathcal{L}^\pm}^{(6)}} {{\mathcal{L}^\pm}'}^2}{ {\mathcal{L}^\pm}^7}-\frac{4407403}{2048}\frac{ {{\mathcal{L}^\pm}''}^3 {{\mathcal{L}^\pm}'}^2}{ {\mathcal{L}^\pm}^9}+\frac{97383 }{128}\frac{{{\mathcal{L}^\pm}''}^2 {{\mathcal{L}^\pm}'''} {{\mathcal{L}^\pm}'}}{ {\mathcal{L}^\pm}^8}\\
	&\hspace{2cm}+\frac{11 }{16}\frac{{{\mathcal{L}^\pm}^{(7)}} {{\mathcal{L}^\pm}'}}{ {\mathcal{L}^\pm}^6}-\frac{1287}{32}\frac{ {{\mathcal{L}^\pm}''} {{\mathcal{L}^\pm}^{(5)}} {{\mathcal{L}^\pm}'}}{ {\mathcal{L}^\pm}^7}-\frac{3861}{64}\frac{ {{\mathcal{L}^\pm}'''} {{\mathcal{L}^\pm}^{(4)}} {{\mathcal{L}^\pm}'}}{ {\mathcal{L}^\pm}^7}\\
	&\hspace{2cm}+\frac{180323}{2048}\frac{ {{\mathcal{L}^\pm}''}^4}{ {\mathcal{L}^\pm}^8}+\frac{253}{128}\frac{ {{\mathcal{L}^\pm}^{(4)}}^2}{ {\mathcal{L}^\pm}^6}+\frac{209}{64}\frac{{{\mathcal{L}^\pm}'''} {{\mathcal{L}^\pm}^{(5)}}}{ {\mathcal{L}^\pm}^6}+\frac{121}{64}\frac{ {{\mathcal{L}^\pm}''} {{\mathcal{L}^\pm}^{(6)}}}{ {\mathcal{L}^\pm}^6}\\
	&\hspace{3.5cm}\left.-\frac{1}{32}\frac{{{\mathcal{L}^\pm}^{(8)}}}{ {\mathcal{L}^\pm}^5}-\frac{13299 }{256 }\frac{{{\mathcal{L}^\pm}''} {{\mathcal{L}^\pm}'''}^2}{{\mathcal{L}^\pm}^7}-\frac{10725}{256}\frac{ {{\mathcal{L}^\pm}''}^2 {{\mathcal{L}^\pm}^{(4)}}}{ {\mathcal{L}^\pm}^7}\right)\,.
	\end{split}
\end{split}
\end{align}
Note that these satisfy the following identity~\cite{tu1989}: 
\begin{align}
	\HHpm_n=\frac{2}{3-2n}\int d\phi\,\frac{\delta H_n^\pm}{\delta \cL^\pm}    \LLpm\,.
\end{align}

\section{Euclidean formulation and the complex Dym equation} \label{euclideanDym}

In this appendix, we construct the Euclidean counterpart of the boundary conditions introduced in \eqref{a} and study the corresponding complex Dym equation. To this end, we follow the Euclidean construction described in \cite{Bunster:2014mua}. 

First, consider a $\mathfrak{sl}(2,\mathbb{C})$ connection $\mathcal A^E$. As in the Lorentzian case, the radial dependence can be removed through a suitable gauge choice,
\begin{equation}
\mathcal A^E=b^{-1}a^E b+b^{-1}d b,
\end{equation}
where $b$ is taken to depend solely on the radial coordinate and to capture the entire radial dependence of the fields. The boundary conditions are then given by
\begin{subequations}\label{EuclideanBC}
\begin{align}
a_\phi^E &= L_{1} - \frac{12\pi}{c} \mathcal{L}_E L_{-1},\label{Euclideanaphi} \\
a_\tau^E &= -\frac{i}{\ell} \left( \mu_E a_\phi^E-\partial_\phi\mu_E L_0 +\frac{1}{2}\partial_\phi^2\mu_E L_{-1} \right) ,\label{Euclideanat}\\
a_\rho^E &= 0.\label{Euclideanarho}
\end{align}
\end{subequations}

Here, the complex function $\mathcal{L}_E$ represents the dynamical mode, while $\mu_E$, also complex, will be chosen as a suitable functional of $\mathcal{L}_E$ and its angular derivatives. Since we restrict the analysis to stationary configurations, both $\mathcal{L}_E$ and $\mu_E$ are taken to depend solely on $\phi$.

The zero-curvature condition reduces to

\begin{equation}
2\mathcal{L}_E \partial_\phi\mu_E+\mu_E \partial_\phi\mathcal{L}_E -\frac{c}{24\pi} \partial_\phi^3\mu_E=0.
\end{equation}
A procedure analogous to that described in \cite{Lara:2024cie} and outlined in the Introduction can then be applied. To recover the first nontrivial member of the Dym hierarchy, we consistently choose
\begin{equation}
\mu_E = 1+\frac{12\pi}{c}\frac{1}{\sqrt{\mathcal{L}_E}}.
\end{equation}
The resulting equation is the stationary complex Dym equation. Its form is identical to that of its real counterpart \eqref{stationary_dym}, except that the dynamical field is now complex-valued,
\begin{equation}\label{EuclideanDym}
\partial_\phi\mathcal{L}_E -\partial_\phi^3 \left( \frac{1}{2\sqrt{\mathcal{L}_E}} \right)=0.
\end{equation}

The complex Dym equation can be integrated twice following the same steps as in Sec. \ref{Periodic Solution}, yielding two complex integration constants, $A_E$ and $B_E$, and fulfilling the analogous relations
\begin{subequations}
 \begin{align}
  \begin{split}\label{Euclideancond1}
    \left(\partial_\phi\LL_E\right)^2&=16{\LL_E}^{5/2}\left(2B_E\sqrt{\LL_E} -\LL_E-A_E\right)\,,
  \end{split}\\
  \begin{split}\label{Euclideancond2}
    \partial_\phi^2 \LL_E& = 4{\LL_E}^{3/2}\left(12B_E\sqrt{\LL_E}-7\LL_E-5A_E\right)\,.
  \end{split}
 \end{align}
\end{subequations}
As explained in Sec. \ref{Euclidean BH}, triviality of the temporal holonomy fixes the Euclidean period $\beta$ to
\begin{equation}
\beta=\sqrt{\frac{c}{12\pi}}
\frac{\ell\pi}
{\sqrt{A_E+\frac{12\pi}{c}\left(2B_E+\frac{12\pi}{c}\right)}}.
\end{equation}
For the configuration to define a regular Euclidean geometry, this quantity must in addition be real and positive.

The Euclidean metric components are then recovered by the usual relation
\begin{equation}
  g_{\mu\nu}^E=\frac{\ell^2}{2}\left< (\mathcal A^E_\mu+{\mathcal A^E_\mu}^\dagger),(\mathcal A^E_\nu+{\mathcal A^E_\nu}^\dagger) \right>,
\end{equation}
resulting in 
\begin{subequations}\label{gE}
  \begin{align} 
    g_{\tau\tau}^E=&\frac{\rho^2}{\ell^2}\left(1+\frac{12\pi}{c\sqrt{\LL_E}}\right)\left(1+\frac{12\pi}{c\sqrt{\overline{\LL_E}}}\right)-\frac{6\pi}{c}\left[2\left(A_E+\overline{A_E}\right)+\frac{48\pi}{c}\left(B_E+\overline{B_E}\right)\right.\nonumber\\
           &\left.+\frac{576\pi^2}{c^2}-\frac{3\pi}{c}\frac{{\LL_E}'{\overline{\LL_E}}'}{{\LL_E}^{3/2}{\overline{\LL_E}}^{3/2}}\right]+\frac{144\pi^2\ell^2}{c^2\rho^2}\left(A_E+\frac{12\pi}{c}\sqrt{\LL_E}\right)\left(\overline{A_E}+\frac{12\pi}{c}\sqrt{\overline{\LL_E}}\right),\label{gttE}\\
    g_{\tau\rho}^E=&-i\frac{3\pi\ell}{c\rho}\left(\frac{{\LL_E}'}{{\LL_E}^{3/2}}-\frac{{\overline{\LL_E}}'}{{\overline{\LL_E}}^{3/2}}\right),\label{gtrhoE}\\
    g_{\tau\phi}^E=&-i \frac{6\pi\rho^2}{c\ell}\left(\frac{1}{\sqrt{\LL_E}}-\frac{1}{\sqrt{\overline{\LL_E}}}\right)-i \frac{6\pi\ell}{c}\left[\left(\LL_E+A_E\right)-\left(\overline{\LL_E}+\overline{A_E}\right)+\frac{24\pi}{c}\left(\sqrt{\LL_E}-\sqrt{\overline{\LL_E}}\right)\right]\nonumber\\
           &-i\frac{72\pi^2\ell^3}{c^2\rho^2}\left[\left(A_E+\frac{12\pi}{c}\sqrt{\LL_E}\right)\overline{\LL_E}-\left(\overline{A_E}+\frac{12\pi}{c}\sqrt{\overline{\LL_E}}\right)\LL_E\right],\label{gtphiE}\\
    g_{\rho\rho}^E=&\frac{\ell^2}{\rho^2},\\
    g_{\phi\phi}^E=&\rho^2 +\frac{12\pi\ell^2}{c}\left(\LL_E+\overline{\LL_E}\right)+\frac{144\pi^2\ell^4}{c^2\rho^2}\LL_E\overline{\LL_E},
  \end{align}
\end{subequations}
where the overline denotes complex conjugation. By construction, the resulting metric is real.

It is useful to note that the Euclidean metric can be formally obtained from the Lorentzian metric \eqref{gmunu} through the identifications
\begin{equation}\label{lortoeuc}
\LL^+ \rightarrow \LL_E,\quad
\LL^- \rightarrow \overline{\LL_E},\quad
A^+ \rightarrow A_E,\quad
A^- \rightarrow \overline{A_E},\quad
B^+ \rightarrow B_E,\quad
B^- \rightarrow \overline{B_E},
\end{equation}
along with  $t \rightarrow i \tau$.

We now seek the complete set of Euclidean solutions that can be connected to the Lorentzian solutions described in Sec. \ref{Metric}. Recall that a Lorentzian solution is characterized by two independent periodic functions, $\LL^+$ and $\LL^-$. Accordingly, its Euclidean counterpart should encode two independent real periods within a single complex solution.

In general, stationary solutions to the complex Dym equation possess two independent periods, although these need not be real. This follows from the fact that periodic solutions are determined by an algebraic curve. Indeed, by introducing
\begin{equation}
\LL_E=\frac{1}{{u_E}^2}, \qquad p_E=u_E\partial_\phi u_E,
\end{equation}
the integrated equation \eqref{Euclideancond1} can be written as
\begin{equation}
p_E^2=-4A_E u_E\left(u_E-u_E^-\right)\left(u_E-u_E^+\right),
\end{equation}
which defines a complex elliptic curve and, consequently, a genus-one Riemann surface.

Finding two real periods therefore amounts to identifying two independent closed cycles on this Riemann surface for which the complex integral
\begin{equation}
T_{\gamma}=\oint_\gamma \frac{u_E\,du_E}{p_E}
=\oint_\gamma
\frac{du_E}
{\sqrt{-\frac{4A_E}{u_E}\left(u_E-u_E^-\right)\left(u_E-u_E^+\right)}},
\end{equation}
takes real values. The elliptic curve possesses four ramification points,
\begin{equation}
\{u_E^+,u_E^-,0,\infty\},
\end{equation}
where $u_E^\pm$ are defined as in \eqref{f}. The associated Riemann surface has the topology of a torus and therefore admits two independent noncontractible cycles. The corresponding periods are, however, generically complex.

Determining the conditions on $A_E$ and $B_E$ for which both periods become real is, in general, a nontrivial problem. Moreover, the formal identification \eqref{lortoeuc} and comparison with the Lorentzian case suggest that the Euclidean solution should not be fully doubly periodic, but encode the two periodicities within different parts of the complex function. This imposes an additional restriction and makes a complete characterization considerably more involved. For this reason, we restrict the analysis to a particular class of complex solutions that can be directly related to the Lorentzian configurations.

A simple class of such solutions is obtained by assuming constant phases,
\begin{equation}
A_E=e^{i\alpha}A,\qquad
B_E=e^{i\gamma}B,\qquad
\LL_E=e^{i\delta}\LL,
\end{equation}
where $A$, $B$, and $\LL$ are real. Within this ansatz, the integrated relation \eqref{Euclideancond1} admits three consistent choices for the phases $(\alpha,\gamma,\delta)$,
\begin{equation}
\left(-\frac{\pi}{3},-\frac{2\pi}{3},\frac{2\pi}{3}\right),\qquad
\left(\frac{\pi}{3},\frac{2\pi}{3},-\frac{2\pi}{3}\right),\qquad
(0,0,0),
\end{equation}
while $\LL$ satisfies the real stationary Dym equation \eqref{cond1}. Hence, every real periodic solution gives rise to three constant-phase solutions of the complex Dym equation.

The two nontrivial branches are related by complex conjugation and therefore give rise to the same Euclidean spacetime \eqref{gE}. However, the corresponding Euclidean period $\beta$ is complex, and hence these branches do not describe a regular real Euclidean geometry with solid-torus topology.

The vanishing-phase scenario remains the only possibility within this family. In this case, the solution is real, $\LL_E=\overline{\LL_E}\equiv \LL$, and the Euclidean metric reduces to
\begin{equation}
ds^2=
\left(\frac{12\pi}{c}\right)^2
\left[
\frac{\rho}{\ell}
\left(
\frac{c}{12\pi}+\frac{1}{\sqrt{\LL}}
\right)
-\frac{\ell}{\rho}
\left(
A+\frac{12\pi}{c}\sqrt{\LL}
\right)
\right]^2 d\tau^2
+\frac{\ell^2}{\rho^2}d\rho^2
+\left(
\rho+\frac{12\pi\ell^2}{c\rho}\LL
\right)^2d\phi^2.
\end{equation}
This is precisely the Euclidean counterpart of the Lorentzian spacetime with vanishing angular momentum described at the end of Sec. \ref{Metric}.

Having constructed the corresponding real and regular Euclidean geometry, the entropy can be evaluated from the Chern-Simons action. In the Euclidean formulation, the entropy is obtained from the boundary term that must be added to the action to make it differentiable. In the Hamiltonian formulation, the variation of such a boundary term is
\begin{equation}
\delta B^E=-\frac{c}{12\pi}\int d\tau d^2x 
\epsilon^{ij}\partial_i\left\langle \mathcal A_\tau^E\delta \mathcal A_j^E\right\rangle .
\end{equation}
This receives contributions from the boundary at infinity and from the horizon. In the present case, the latter is described by
\begin{equation}
\rho=\rho_H(\phi).
\end{equation}
Since the radial coordinate ranges from $\rho_H(\phi)$ to infinity, Stokes's theorem gives
\begin{equation}
\delta B^E=-\frac{c}{12\pi}
\left.\int d\tau d\phi 
\left\langle \mathcal A_\tau^E\delta \mathcal A_\phi^E\right\rangle
\right|_{\rho=\infty}\
+\frac{c}{12\pi}
\left.\int d\tau d\phi 
\left\langle \mathcal A_\tau^E
\left(\delta \mathcal A_\phi^E+\rho_H'\delta \mathcal A_\rho^E\right)
\right\rangle\right|_{\rho=\rho_H(\phi)} .
\end{equation}
The first term contains the contributions from the asymptotic charges, while the second gives the horizon contribution. The horizon corresponds to the hypersurface $\rho=\rho_H(\phi)$, and therefore $d\rho=\rho_H'(\phi)\,d\phi$. Hence, restricting the connection $\mathcal A^E=\mathcal A_\tau^E d\tau+\mathcal A_\rho^E d\rho+\mathcal A_\phi^E d\phi$ to the horizon gives
\begin{equation}
\left. \mathcal A^E \right|_{\rho=\rho_H}
=
\mathcal A_\tau^E d\tau
+
\left(\mathcal A_\phi^E+\rho_H'\mathcal A_\rho^E\right)d\phi .
\end{equation}
Thus, the combination appearing in the horizon term is simply the component of the connection tangent to the horizon.

Equivalently, one may introduce the horizon-adapted radial coordinate
\begin{equation}
r=\rho-\rho_H(\phi),
\end{equation}
for which the horizon is located at the fixed position $r=0$. In this coordinate system, the variation is performed keeping the position of the inner boundary fixed, so that $\rho_H(\phi)$ is not independently varied.

On the other hand, Euclidean regularity fixes the holonomy of the connection along the contractible temporal cycle. Using a Euclidean-time coordinate with fixed period, and choosing a gauge in which a fixed representative of this holonomy is kept fixed at the horizon \cite{Bunster:2014mua}, one may impose
\begin{equation}
\left.\delta \mathcal A_\tau^E\right|_{\rho=\rho_H}=0 .
\end{equation}
The horizon contribution is then integrable. Recalling that the Euclidean Einstein-Hilbert action is recovered from the $\mathfrak{sl}(2,\mathbb{C})$ Chern-Simons action as $I_{EH}=i(I_{CS}[\mathcal A^E]-I_{CS}[{\mathcal A^E}^\dagger])$, then the entropy is given by

\begin{equation}
S
=
-\frac{c}{6\pi}\Im
\left.
\int d\tau\,d\phi\,
\left\langle
\mathcal A_\tau^E
\left(
\mathcal A_\phi^E+\rho_H'\mathcal A_\rho^E
\right)
\right\rangle
\right|_{\rho=\rho_H(\phi)} .
\end{equation}

Using the horizon defined in \eqref{horizon}, this gives
\begin{equation}\label{entropy}
S=2\pi \left( \frac{c}{12\pi} \right)^{3/2}\sqrt{A+\frac{12\pi}{c}\left(2B+\frac{12\pi}{c} \right)}\int_0^{2\pi}d\phi\frac{\left(A+\frac{24 \pi}{c}\sqrt{\LL}+\LL \right)\sqrt{\LL}}{\left(1+\frac{c}{12\pi}\sqrt{\LL}\right)\left(A+\frac{12\pi}{c}\sqrt{\LL}\right)}
\end{equation}
which agrees with one quarter of the horizon length \eqref{horizonarea}, as expected from the Bekenstein-Hawking formula.

\section{Integration of the $N$-th Stationary Dym Equation} \label{proof}

The objective of this appendix is to demonstrate that Eq. \eqref{DymNintegrada} becomes a total derivative and can be readily integrated when multiplied by appropriate terms. In what follows, we drop the $\pm$ signs. We begin by multiplying by the term $\partial_\phi \left( \frac{\delta H_M}{\delta \cL}\right)$, yielding, 
 \begin{equation}\label{appendix1}
    \cL \partial_\phi \left( \frac{\delta H_M}{\delta \cL}\right) - \partial_\phi^2 \left(\frac{\delta H_N}{\delta \cL}\right) \partial_\phi \left( \frac{\delta H_M}{\delta \cL}\right) = A \partial_\phi \left( \frac{\delta H_M}{\delta \cL}\right).
\end{equation}

First, the term on the right-hand side of the equation is a total derivative since $A$ is a constant.

Next, consider the recursion relation \eqref{recursion}, which, after some straightforward manipulation, can be expressed as
\begin{align}
  \cL \partial_\phi \left( \frac{\delta H_{n+1}}{\delta \cL} \right) = \partial_\phi \left[ \partial_\phi^2 \left( \frac{\delta H_n}{\delta \cL}\right) - \cL \left( \frac{\delta H_{n+1}}{\delta \cL} \right) \right].
\end{align}
Thus, the first term on the left-hand side of Eq. \eqref{appendix1} also yields a total derivative.

The proof that the second term on the left-hand side is a total derivative follows the same logic as demonstrating that the Dym conserved quantities $H_n$ span an Abelian algebra under both Poisson brackets defined by the operators $\cD$ and $\cE$ (see~\cite{olver2000applications}). First, note that both operators are antisymmetric, in the sense that
\begin{align}\label{antiD}
  f \cD g + g \cD f &= \partial_\phi \left( 2 \cL f g \right), \\
  f \cE g + g \cE f &= \partial_\phi \left( f \partial_\phi^2 g - \partial_\phi f \partial_\phi g + g \partial_\phi^2 f \right),\label{antiE}
\end{align}
for any functions $f$ and $g$. These results, along with the recursion relation \eqref{recursion}, lead to the identity
\begin{align}
 \cE \left( \frac{\delta H_N}{\delta \cL} \right) \left( \frac{\delta H_M}{\delta \cL} \right) = \cE \left( \frac{\delta H_{N+1}}{\delta \cL} \right) \left( \frac{\delta H_{M-1}}{\delta \cL} \right) + \text{boundary terms}.
\end{align}
Assuming $M>N$, without loss of generality, this identity is applied recursively $M-N$ times, implying
\begin{align}
 \cE \left( \frac{\delta H_N}{\delta \cL} \right) \left( \frac{\delta H_M}{\delta \cL} \right) = \cE \left( \frac{\delta H_{M}}{\delta \cL} \right) \left( \frac{\delta H_{N}}{\delta \cL} \right) + \text{boundary terms}.
\end{align}
Combining this result with the antisymmetry relation \eqref{antiE}, it follows that $\cE \left( \frac{\delta H_N}{\delta \cL} \right) \left( \frac{\delta H_M}{\delta \cL} \right)$ is a boundary term. Since the second term on the left-hand side of equation \eqref{appendix1} differs only by a boundary term, it follows that this term is also a boundary term. Thus, the complete Eq. \eqref{appendix1} is a total derivative and can be readily integrated.

\clearpage
\printbibliography

@article{Brown:1986nw,
	author = "Brown, J. David and Henneaux, M.",
	title = "{Central Charges in the Canonical Realization of Asymptotic Symmetries: An Example from Three-Dimensional Gravity}",
	doi = "10.1007/BF01211590",
	journal = "Commun. Math. Phys.",
	volume = "104",
	pages = "207--226",
	year = "1986"
}

@article{Regge:1974zd,
	author = "Regge, Tullio and Teitelboim, Claudio",
	title = "{Role of Surface Integrals in the Hamiltonian Formulation of General Relativity}",
	reportNumber = "Print-74-0988 (IAS,PRINCETON)",
	doi = "10.1016/0003-4916(74)90404-7",
	journal = "Annals Phys.",
	volume = "88",
	pages = "286",
	year = "1974"
}

@article{Giribet:2023xom,
    author = "Giribet, Gaston and Laurnagaray, Juan and Schmied, Pedro",
    title = "{Probing the near-horizon geometry of black rings}",
    eprint = "2304.14461",
    archivePrefix = "arXiv",
    primaryClass = "hep-th",
    doi = "10.1103/PhysRevD.108.024061",
    journal = "Phys. Rev. D",
    volume = "108",
    number = "2",
    pages = "024061",
    year = "2023"
}

@article{Giribet:2021zie,
    author = "Giribet, Gaston and Montecchio, Luciano",
    title = "{Colored black holes and Kac-Moody algebra}",
    eprint = "2111.08178",
    archivePrefix = "arXiv",
    primaryClass = "hep-th",
    doi = "10.1103/PhysRevD.105.064006",
    journal = "Phys. Rev. D",
    volume = "105",
    number = "6",
    pages = "064006",
    year = "2022"
}

@article{Barnich:2010eb,
    author = "Barnich, Glenn and Troessaert, Cedric",
    title = "{Aspects of the BMS/CFT correspondence}",
    eprint = "1001.1541",
    archivePrefix = "arXiv",
    primaryClass = "hep-th",
    reportNumber = "ULB-TH-09-28",
    doi = "10.1007/JHEP05(2010)062",
    journal = "JHEP",
    volume = "05",
    pages = "062",
    year = "2010"
}

@article{Donnay:2015abr,
    author = "Donnay, Laura and Giribet, Gaston and Gonzalez, Hernan A. and Pino, Miguel",
    title = "{Supertranslations and Superrotations at the Black Hole Horizon}",
    eprint = "1511.08687",
    archivePrefix = "arXiv",
    primaryClass = "hep-th",
    doi = "10.1103/PhysRevLett.116.091101",
    journal = "Phys. Rev. Lett.",
    volume = "116",
    number = "9",
    pages = "091101",
    year = "2016"
}

@article{Donnay:2016ejv,
    author = "Donnay, Laura and Giribet, Gaston and Gonz\'alez, Hern\'an A. and Pino, Miguel",
    title = "{Extended Symmetries at the Black Hole Horizon}",
    eprint = "1607.05703",
    archivePrefix = "arXiv",
    primaryClass = "hep-th",
    doi = "10.1007/JHEP09(2016)100",
    journal = "JHEP",
    volume = "09",
    pages = "100",
    year = "2016"
}

@article{Kehagias:2016zry,
	author = "Kehagias, Alex and Riotto, Antonio",
	title = "{BMS in Cosmology}",
	eprint = "1602.02653",
	archivePrefix = "arXiv",
	primaryClass = "hep-th",
	doi = "10.1088/1475-7516/2016/05/059",
	journal = "JCAP",
	volume = "05",
	pages = "059",
	year = "2016"
}

@article{Bonga:2020fhx,
	author = "Bonga, B\'eatrice and Prabhu, Kartik",
	title = "{BMS-like symmetries in cosmology}",
	eprint = "2009.01243",
	archivePrefix = "arXiv",
	primaryClass = "gr-qc",
	doi = "10.1103/PhysRevD.102.104043",
	journal = "Phys. Rev. D",
	volume = "102",
	number = "10",
	pages = "104043",
	year = "2020"
}

@article{Enriquez-Rojo:2020miw,
	author = "Enriquez-Rojo, Martin and Heckelbacher, Till",
	title = "{Asymptotic symmetries in spatially flat FRW spacetimes}",
	eprint = "2011.01960",
	archivePrefix = "arXiv",
	primaryClass = "hep-th",
	reportNumber = "LMU-ASC 42/20",
	doi = "10.1103/PhysRevD.103.064009",
	journal = "Phys. Rev. D",
	volume = "103",
	number = "6",
	pages = "064009",
	year = "2021"
}

@article{Enriquez-Rojo:2021blc,
	author = "Enriquez-Rojo, Martin and Heckelbacher, Till",
	title = "{Holography and black holes in asymptotically flat FLRW spacetimes}",
	eprint = "2102.02234",
	archivePrefix = "arXiv",
	primaryClass = "gr-qc",
	reportNumber = "LMU-ASC 03/21",
	doi = "10.1103/PhysRevD.103.104035",
	journal = "Phys. Rev. D",
	volume = "103",
	number = "10",
	pages = "104035",
	year = "2021"
}

@article{Troessaert:2017jcm,
	author = "Troessaert, C\'edric",
	title = "{The BMS4 algebra at spatial infinity}",
	eprint = "1704.06223",
	archivePrefix = "arXiv",
	primaryClass = "hep-th",
	doi = "10.1088/1361-6382/aaae22",
	journal = "Class. Quant. Grav.",
	volume = "35",
	number = "7",
	pages = "074003",
	year = "2018"
}

@article{Fuentealba:2020zkf,
	author = "Fuentealba, Oscar and Gonz\'alez, Hern\'an A. and P\'erez, Alfredo and Tempo, David and Troncoso, Ricardo",
	title = "{Superconformal Bondi-Metzner-Sachs Algebra in Three Dimensions}",
	eprint = "2011.08197",
	archivePrefix = "arXiv",
	primaryClass = "hep-th",
	reportNumber = "CECS-PHY-20/04",
	doi = "10.1103/PhysRevLett.126.091602",
	journal = "Phys. Rev. Lett.",
	volume = "126",
	number = "9",
	pages = "091602",
	year = "2021"
}

@article{Guica:2008mu,
	author = "Guica, Monica and Hartman, Thomas and Song, Wei and Strominger, Andrew",
	title = "{The Kerr/CFT Correspondence}",
	eprint = "0809.4266",
	archivePrefix = "arXiv",
	primaryClass = "hep-th",
	doi = "10.1103/PhysRevD.80.124008",
	journal = "Phys. Rev. D",
	volume = "80",
	pages = "124008",
	year = "2009"
}

@article{Henneaux:2002wm,
	author = "Henneaux, Marc and Martinez, Cristian and Troncoso, Ricardo and Zanelli, Jorge",
	title = "{Black holes and asymptotics of 2+1 gravity coupled to a scalar field}",
	eprint = "hep-th/0201170",
	archivePrefix = "arXiv",
	reportNumber = "CECS-PHY-01-09, ULB-TH-02-07",
	doi = "10.1103/PhysRevD.65.104007",
	journal = "Phys. Rev. D",
	volume = "65",
	pages = "104007",
	year = "2002"
}

@article{Henneaux:2011hv,
	author = "Henneaux, Marc and Martinez, Cristian and Troncoso, Ricardo",
	title = "{Asymptotically warped anti-de Sitter spacetimes in topologically massive gravity}",
	eprint = "1108.2841",
	archivePrefix = "arXiv",
	primaryClass = "hep-th",
	reportNumber = "CECS-PHY-11-05",
	doi = "10.1103/PhysRevD.84.124016",
	journal = "Phys. Rev. D",
	volume = "84",
	pages = "124016",
	year = "2011"
}

@article{Henneaux:2010fy,
	author = "Henneaux, Marc and Martinez, Cristian and Troncoso, Ricardo",
	title = "{More on Asymptotically Anti-de Sitter Spaces in Topologically Massive Gravity}",
	eprint = "1006.0273",
	archivePrefix = "arXiv",
	primaryClass = "hep-th",
	reportNumber = "CECS-PHY-10-07",
	doi = "10.1103/PhysRevD.82.064038",
	journal = "Phys. Rev. D",
	volume = "82",
	pages = "064038",
	year = "2010"
}

@article{Liu:2009kc,
	author = "Liu, Yan and Sun, Ya-Wen",
	title = "{Consistent Boundary Conditions for New Massive Gravity in $AdS_3$}",
	eprint = "0903.2933",
	archivePrefix = "arXiv",
	primaryClass = "hep-th",
	doi = "10.1088/1126-6708/2009/05/039",
	journal = "JHEP",
	volume = "05",
	pages = "039",
	year = "2009"
}

@article{Compere:2013bya,
	author = "Comp\`ere, Geoffrey and Song, Wei and Strominger, Andrew",
	title = "{New Boundary Conditions for AdS3}",
	eprint = "1303.2662",
	archivePrefix = "arXiv",
	primaryClass = "hep-th",
	doi = "10.1007/JHEP05(2013)152",
	journal = "JHEP",
	volume = "05",
	pages = "152",
	year = "2013"
}

@article{Afshar:2015wjm,
	author = "Afshar, Hamid and Detournay, St\'ephane and Grumiller, Daniel and Oblak, Blagoje",
	title = "{Near-Horizon Geometry and Warped Conformal Symmetry}",
	eprint = "1512.08233",
	archivePrefix = "arXiv",
	primaryClass = "hep-th",
	reportNumber = "TUW-15-22",
	doi = "10.1007/JHEP03(2016)187",
	journal = "JHEP",
	volume = "03",
	pages = "187",
	year = "2016"
}

@article{Troessaert:2013fma,
	author = "Troessaert, C\'edric",
	title = "{Enhanced asymptotic symmetry algebra of $AdS_{3}$}",
	eprint = "1303.3296",
	archivePrefix = "arXiv",
	primaryClass = "hep-th",
	doi = "10.1007/JHEP08(2013)044",
	journal = "JHEP",
	volume = "08",
	pages = "044",
	year = "2013"
}

@article{Alessio:2020ioh,
	author = "Alessio, Francesco and Barnich, Glenn and Ciambelli, Luca and Mao, Pujian and Ruzziconi, Romain",
	title = "{Weyl charges in asymptotically locally AdS$_3$ spacetimes}",
	eprint = "2010.15452",
	archivePrefix = "arXiv",
	primaryClass = "hep-th",
	reportNumber = "CJQS-2021-002",
	doi = "10.1103/PhysRevD.103.046003",
	journal = "Phys. Rev. D",
	volume = "103",
	number = "4",
	pages = "046003",
	year = "2021"
}

@article{Ozer:2019nkv,
    author = {\"Ozer, H. T. and Filiz, Ayt\"ul},
    title = "{Exploring new boundary conditions for $\mathcal {N}=(1,1)$ extended higher-spin $AdS_3$ supergravity}",
    eprint = "1907.06104",
    archivePrefix = "arXiv",
    primaryClass = "hep-th",
    doi = "10.1140/epjc/s10052-020-08613-4",
    journal = "Eur. Phys. J. C",
    volume = "80",
    number = "11",
    pages = "1072",
    year = "2020"
}

@article{Ozer:2021wtx,
    author = {\"Ozer, H. T. and Filiz, Ayt\"ul},
    title = "{${\mathcal {N}}=(2,2)$ extended ${\mathfrak {sl}}(3|2)$ Chern\textendash{}Simons $AdS_3$ supergravity with new boundaries}",
    eprint = "2107.11069",
    archivePrefix = "arXiv",
    primaryClass = "hep-th",
    doi = "10.1140/epjc/s10052-022-10422-w",
    journal = "Eur. Phys. J. C",
    volume = "82",
    number = "5",
    pages = "472",
    year = "2022"
}

@article{Avery:2013dja,
	author = "Avery, Steven G. and Poojary, Rohan R. and Suryanarayana, Nemani V.",
	title = "{An sl(2,$\mathbb{R}$) current algebra from $AdS_3$ gravity}",
	eprint = "1304.4252",
	archivePrefix = "arXiv",
	primaryClass = "hep-th",
	doi = "10.1007/JHEP01(2014)144",
	journal = "JHEP",
	volume = "01",
	pages = "144",
	year = "2014"
}

@article{Grumiller:2016pqb,
	author = "Grumiller, Daniel and Riegler, Max",
	title = "{Most general AdS$_{3}$ boundary conditions}",
	eprint = "1608.01308",
	archivePrefix = "arXiv",
	primaryClass = "hep-th",
	reportNumber = "TUW-16-16",
	doi = "10.1007/JHEP10(2016)023",
	journal = "JHEP",
	volume = "10",
	pages = "023",
	year = "2016"
}

@article{Valcarcel:2018kwd,
	author = "Valc\'arcel, C. E.",
	title = "{New boundary conditions for (extended) $\mathrm{AdS}_3$ supergravity}",
	eprint = "1812.02799",
	archivePrefix = "arXiv",
	primaryClass = "hep-th",
	doi = "10.1088/1361-6382/ab04da",
	journal = "Class. Quant. Grav.",
	volume = "36",
	pages = "065002",
	year = "2019"
}

@article{Henneaux:2019sjx,
	author = "Henneaux, Marc and Merbis, Wout and Ranjbar, Arash",
	title = "{Asymptotic dynamics of AdS$_3$ gravity with two asymptotic regions}",
	eprint = "1912.09465",
	archivePrefix = "arXiv",
	primaryClass = "hep-th",
	doi = "10.1007/JHEP03(2020)064",
	journal = "JHEP",
	volume = "03",
	pages = "064",
	year = "2020"
}

@article{Prema:2021sjp,
    author = "Prema, Aneesh B. and Comp\`ere, Geoffrey and Pipolo de Gioia, Leonardo and Mol, Igor and Swidler, Bianca",
    title = "{Celestial holography: Lectures on asymptotic symmetries}",
    eprint = "2109.00997",
    archivePrefix = "arXiv",
    primaryClass = "hep-th",
    doi = "10.21468/SciPostPhysLectNotes.47",
    journal = "SciPost Phys. Lect. Notes",
    volume = "47",
    pages = "1",
    year = "2022"
}

@article{Pasterski:2023ikd,
	author = "Pasterski, Sabrina",
	title = "{A Chapter on Celestial Holography}",
	eprint = "2310.04932",
	archivePrefix = "arXiv",
	primaryClass = "hep-th",
	month = "10",
	year = "2023"
}

@article{Hawking:2016sgy,
    author = "Hawking, Stephen W. and Perry, Malcolm J. and Strominger, Andrew",
    title = "{Superrotation Charge and Supertranslation Hair on Black Holes}",
    eprint = "1611.09175",
    archivePrefix = "arXiv",
    primaryClass = "hep-th",
    doi = "10.1007/JHEP05(2017)161",
    journal = "JHEP",
    volume = "05",
    pages = "161",
    year = "2017"
}

@article{Son:2008ye,
	author = "Son, D. T.",
	title = "{Toward an AdS/cold atoms correspondence: A Geometric realization of the Schrodinger symmetry}",
	eprint = "0804.3972",
	archivePrefix = "arXiv",
	primaryClass = "hep-th",
	reportNumber = "INT-PUB-08-08",
	doi = "10.1103/PhysRevD.78.046003",
	journal = "Phys. Rev. D",
	volume = "78",
	pages = "046003",
	year = "2008"
}

@article{Achucarro:1986uwr,
    author = "Achucarro, A. and Townsend, P. K.",
    editor = "Salam, A. and Sezgin, E.",
    title = "{A Chern-Simons Action for Three-Dimensional anti-De Sitter Supergravity Theories}",
    reportNumber = "Print-87-0078 (CAMBRIDGE)",
    doi = "10.1016/0370-2693(86)90140-1",
    journal = "Phys. Lett. B",
    volume = "180",
    pages = "89",
    year = "1986"
}

@article{Witten:1988hc,
    author = "Witten, Edward",
    title = "{(2+1)-Dimensional Gravity as an Exactly Soluble System}",
    reportNumber = "IASSNS-HEP-88-32",
    doi = "10.1016/0550-3213(88)90143-5",
    journal = "Nucl. Phys. B",
    volume = "311",
    pages = "46",
    year = "1988"
}

@article{Coussaert:1995zp,
	author = "Coussaert, Oliver and Henneaux, Marc and van Driel, Peter",
	title = "{The Asymptotic dynamics of three-dimensional Einstein gravity with a negative cosmological constant}",
	eprint = "gr-qc/9506019",
	archivePrefix = "arXiv",
	reportNumber = "ULB-TH-95-08",
	doi = "10.1088/0264-9381/12/12/012",
	journal = "Class. Quant. Grav.",
	volume = "12",
	pages = "2961--2966",
	year = "1995"
}

@article{Bunster:2014mua,
    author = "Bunster, Claudio and Henneaux, Marc and Perez, Alfredo and Tempo, David and Troncoso, Ricardo",
    title = "{Generalized Black Holes in Three-dimensional Spacetime}",
    eprint = "1404.3305",
    archivePrefix = "arXiv",
    primaryClass = "hep-th",
    doi = "10.1007/JHEP05(2014)031",
    journal = "JHEP",
    volume = "05",
    pages = "031",
    year = "2014"
}

@article{Afshar:2016wfy,
    author = "Afshar, Hamid and Detournay, Stephane and Grumiller, Daniel and Merbis, Wout and Perez, Alfredo and Tempo, David and Troncoso, Ricardo",
    title = "{Soft Heisenberg hair on black holes in three dimensions}",
    eprint = "1603.04824",
    archivePrefix = "arXiv",
    primaryClass = "hep-th",
    reportNumber = "TUW-16-06",
    doi = "10.1103/PhysRevD.93.101503",
    journal = "Phys. Rev. D",
    volume = "93",
    number = "10",
    pages = "101503",
    year = "2016"
}

@article{PhysRevD.51.622,
	title = {Aspects of black hole quantum mechanics and thermodynamics in 2+1 dimensions},
	author = {Carlip, Steven and Teitelboim, Claudio},
	journal = {Phys. Rev. D},
	volume = {51},
	issue = {2},
	pages = {622--631},
	numpages = {0},
	year = {1995},
	month = {Jan},
	publisher = {American Physical Society},
	doi = {10.1103/PhysRevD.51.622},
	url = {https://link.aps.org/doi/10.1103/PhysRevD.51.622}
}

@article{Martinec:1998wm,
	author = "Martinec, Emil J.",
	title = "{Conformal field theory, geometry, and entropy}",
	eprint = "hep-th/9809021",
	archivePrefix = "arXiv",
	reportNumber = "EFI-98-40",
	month = "9",
	year = "1998"
}

@article{Perez:2016vqo,
    author = "P\'erez, Alfredo and Tempo, David and Troncoso, Ricardo",
    title = "{Boundary conditions for General Relativity on AdS$_{3}$ and the KdV hierarchy}",
    eprint = "1605.04490",
    archivePrefix = "arXiv",
    primaryClass = "hep-th",
    reportNumber = "CECS-PHY-16-04",
    doi = "10.1007/JHEP06(2016)103",
    journal = "JHEP",
    volume = "06",
    pages = "103",
    year = "2016"
}

@article{Dymarsky:2020tjh,
	author = "Dymarsky, Anatoly and Sugishita, Sotaro",
	title = "{KdV-charged black holes}",
	eprint = "2002.08368",
	archivePrefix = "arXiv",
	primaryClass = "hep-th",
	doi = "10.1007/JHEP05(2020)041",
	journal = "JHEP",
	volume = "05",
	pages = "041",
	year = "2020"
}

@article{Ojeda:2020bgz,
    author = "Ojeda, Emilio and P\'erez, Alfredo",
    title = "{Integrable systems and the boundary dynamics of higher spin gravity on AdS$_{3}$}",
    eprint = "2009.07829",
    archivePrefix = "arXiv",
    primaryClass = "hep-th",
    reportNumber = "CECS-PHY-20/02",
    doi = "10.1007/JHEP11(2020)089",
    journal = "JHEP",
    volume = "11",
    pages = "089",
    year = "2020"
}

@article{Ojeda:2019xih,
    author = "Ojeda, Emilio and P\'erez, Alfredo",
    title = "{Boundary conditions for General Relativity in three-dimensional spacetimes, integrable systems and the KdV/mKdV hierarchies}",
    eprint = "1906.11226",
    archivePrefix = "arXiv",
    primaryClass = "hep-th",
    reportNumber = "CECS-PHY-19/02",
    doi = "10.1007/JHEP08(2019)079",
    journal = "JHEP",
    volume = "08",
    pages = "079",
    year = "2019"
}

@article{Fuentealba:2017omf,
    author = "Fuentealba, Oscar and Matulich, Javier and P\'erez, Alfredo and Pino, Miguel and Rodr\'\i{}guez, Pablo and Tempo, David and Troncoso, Ricardo",
    title = "{Integrable systems with BMS$_{3}$ Poisson structure and the dynamics of locally flat spacetimes}",
    eprint = "1711.02646",
    archivePrefix = "arXiv",
    primaryClass = "hep-th",
    reportNumber = "CECS-PHY-17-02",
    doi = "10.1007/JHEP01(2018)148",
    journal = "JHEP",
    volume = "01",
    pages = "148",
    year = "2018"
}

@article{Cardenas:2021vwo,
    author = "C\'ardenas, Marcela and Correa, Francisco and Lara, Kristiansen and Pino, Miguel",
    title = "{Integrable Systems and Spacetime Dynamics}",
    eprint = "2104.09676",
    archivePrefix = "arXiv",
    primaryClass = "hep-th",
    doi = "10.1103/PhysRevLett.127.161601",
    journal = "Phys. Rev. Lett.",
    volume = "127",
    number = "16",
    pages = "161601",
    year = "2021"
}

@article{Cardenas:2024hah,
	author = "C{\'a}rdenas, Marcela",
	title = "{KdV conformal symmetry breaking in nearly AdS$_{2}$}",
	eprint = "2405.03128",
	archivePrefix = "arXiv",
	primaryClass = "hep-th",
	doi = "10.1007/JHEP10(2024)052",
	journal = "JHEP",
	volume = "10",
	pages = "052",
	year = "2024"
}

@article{Cardenas:2025qqi,
    author = "C{\'a}rdenas, Marcela and Correa, Francisco and Pino, Miguel",
    title = "{Integrable black hole dynamics in the asymptotic structure of AdS3}",
    eprint = "2504.20292",
    archivePrefix = "arXiv",
    primaryClass = "hep-th",
    doi = "10.1103/n2hm-lk94",
    journal = "Phys. Rev. D",
    volume = "114",
    number = "2",
    pages = "026010",
    year = "2026"
}

@article{Gonzalez:2018jgp,
	author = "Gonz\'alez, Hern\'an A. and Matulich, Javier and Pino, Miguel and Troncoso, Ricardo",
	title = "{Revisiting the asymptotic dynamics of General Relativity on AdS$_{3}$}",
	eprint = "1809.02749",
	archivePrefix = "arXiv",
	primaryClass = "hep-th",
	reportNumber = "CECS-PHY-18/03",
	doi = "10.1007/JHEP12(2018)115",
	journal = "JHEP",
	volume = "12",
	pages = "115",
	year = "2018"
}

@Inbook{Kruskal1975,
author="Kruskal, Martin",
editor="Moser, J.",
title="Nonlinear wave equations",
bookTitle="Dynamical Systems, Theory and Applications: Battelle Seattle 1974 Rencontres",
year="1975",
publisher="Springer Berlin Heidelberg",
address="Berlin, Heidelberg",
pages="310--354",
isbn="978-3-540-37505-0",
doi="10.1007/3-540-07171-7_9",
url="https://doi.org/10.1007/3-540-07171-7_9"
}

@book{olver2000applications,
  title={Applications of Lie Groups to Differential Equations},
  author={Olver, P.J.},
  isbn={9780387950006},
  lccn={92044573},
  series={Graduate Texts in Mathematics},
  url={https://books.google.cl/books?id=sI2bAxgLMXYC},
  year={2000},
  publisher={Springer New York}
}

@article{tu1989,
	author = {Tu,Gui‐zhang },
	title = {The trace identity, a powerful tool for constructing the Hamiltonian structure of integrable systems},
	journal = {Journal of Mathematical Physics},
	volume = {30},
	number = {2},
	pages = {330-338},
	year = {1989},
	doi = {10.1063/1.528449},
	
	URL = { 
	https://doi.org/10.1063/1.528449
	
	},
	eprint = { 
	https://doi.org/10.1063/1.528449
	
	}
	
}

@article{SolitonsHD,
	author = {Wadati, Miki and Ichikawa, Yoshi H. and Shimizu, Toru},
	title = "{Cusp Soliton of a New Integrable Nonlinear Evolution Equation}",
	journal = {Progress of Theoretical Physics},
	volume = {64},
	number = {6},
	pages = {1959-1967},
	year = {1980},
	month = {12},
	issn = {0033-068X},
	doi = {10.1143/PTP.64.1959},
	url = {https://doi.org/10.1143/PTP.64.1959},
	eprint = {https://academic.oup.com/ptp/article-pdf/64/6/1959/5272896/64-6-1959.pdf},
}

@Article{Novikov1999,
	author = {Novikov, D. P.},
	title = {Algebraic-geometric solutions of the Harry Dym equation},
	journal = {Siberian Mathematical Journal},
	year = {1999},
	volume = {40},
	number = {1},
	pages = {136--140},
	issn = {1573-9260},
	doi = {10.1007/BF02674299},
	url = {https://doi.org/10.1007/BF02674299},
}

@Article{Mokhtari2011,
	author = {Mokhtari, Reza},
	title = {Exact solutions of the Harry Dym equation},
	journal = {Communications in Theoretical Physics},
	year = {2011},
	volume = {55},
	number = {2},
	pages = {204--208},
	month = {Feb},
}

@Article{RoyChowdhuryMukherjee1984,
	author = {A. Roy Chowdhury and Rebecca Mukherjee},
	title = {Elliptic Solutions, Recursion Operators and Complete Lie-Backlund Symmetry for the Harry-Dym Equation},
	journal = {Physica Scripta},
	year = {1984},
	volume = {29},
	number = {4},
	pages = {293},
	doi = {10.1088/0031-8949/29/4/002},
	note = {Published under licence by IOP Publishing Ltd},
}

@article{Magri1978,
	author = {Franco Magri},
	title = {A simple model of the integrable Hamiltonian equation},
	journal = {J. Math. Phys.},
	year = {1978},
	volume = {19},
	pages = {1156--1162},
	doi = {10.1063/1.523777},
	url = {https://doi.org/10.1063/1.523777},
}

@article{Fuchssteiner_1992,
	doi = {10.1088/0305-4470/25/1/025},
	url = {https://dx.doi.org/10.1088/0305-4470/25/1/025},
	year = {1992},
	month = {jan},
	publisher = {},
	volume = {25},
	number = {1},
	pages = {223},
	author = {B Fuchssteiner and  T Schulze and  S Carillo},
	title = {Explicit solutions for the Harry Dym equation},
	journal = {Journal of Physics A: Mathematical and General}
}

@article{Bordag:1995dp,
	author = "Bordag, L. A.",
	title = "{Periodical solutions of the Harry Dym equation and related problems}",
	reportNumber = "LEIPZIG-16-95",
	month = "6",
	year = "1995"
}

@article{alezGaxiola2018ITERATIVEMF,
    author  = {Gonz{\'a}lez-Gaxiola, O. and Ruiz de Ch{\'a}vez, J. and Edeki, S. O.},
    title   = {Iterative method for constructing analytical solutions to the Harry-Dym initial value problem},
    journal = {International Journal of Applied Mathematics},
    volume  = {31},
    number  = {4},
    pages   = {627--640},
    year    = {2018},
    doi     = {10.12732/ijam.v31i4.8}
}

@article{XIAO2019123248,
	title = {Long time behavior and soliton solution for the Harry Dym equation},
	journal = {Journal of Mathematical Analysis and Applications},
	volume = {480},
	number = {1},
	pages = {123248},
	year = {2019},
	issn = {0022-247X},
	doi = {https://doi.org/10.1016/j.jmaa.2019.06.019},
	url = {https://www.sciencedirect.com/science/article/pii/S0022247X19304937},
	author = {Yu Xiao and Engui Fan}
}

@article{Assabaai2022ExactSO,
    author  = {Assabaai, M. A. and Mukherij, O. F.},
    title   = {Exact solutions of the Harry Dym Equation using Lie group method},
    journal = {University of Aden Journal of Natural and Applied Sciences},
    volume  = {24},
    number  = {2},
    pages   = {481--487},
    year    = {2020},
    doi     = {10.47372/uajnas.2020.n2.a15}
}

@article{LI2020106276,
	title = {On the double-pole and two-soliton solutions of the Harry Dym equation},
	journal = {Applied Mathematics Letters},
	volume = {104},
	pages = {106276},
	year = {2020},
	issn = {0893-9659},
	doi = {https://doi.org/10.1016/j.aml.2020.106276},
	url = {https://www.sciencedirect.com/science/article/pii/S0893965920300690},
	author = {Hongyu Li and Jian Xu}
}

@article{Li2017,
	title = {Algebro-Geometric Solutions of the Harry Dym Hierarchy},
	url = {https://doi.org/10.1515/ijnsns-2016-0057},
	author = {Zhu Li},
	pages = {129--136},
	volume = {18},
	number = {2},
	journal = {International Journal of Nonlinear Sciences and Numerical Simulation},
	doi = {doi:10.1515/ijnsns-2016-0057},
	year = {2017},
	lastchecked = {2024-01-17}
}

@article{Banados:1998gg,
	author = "Banados, Maximo",
	editor = "Falomir, H. and Gamboa Saravi, R. E. and Schaposnik, F. A.",
	title = "{Three-dimensional quantum geometry and black holes}",
	eprint = "hep-th/9901148",
	archivePrefix = "arXiv",
	doi = "10.1063/1.59661",
	journal = "AIP Conf. Proc.",
	volume = "484",
	number = "1",
	pages = "147--169",
	year = "1999"
}

@article{Lara:2024cie,
	author = "Lara, Kristiansen and Pino, Miguel and Reyes, Francisco",
	title = "{1/c deformations of AdS$_{3}$ boundary conditions and the Dym hierarchy}",
	eprint = "2401.12338",
	archivePrefix = "arXiv",
	primaryClass = "hep-th",
	doi = "10.1007/JHEP11(2024)042",
	journal = "JHEP",
	volume = "11",
	pages = "042",
	year = "2024"
}

@article{PhysRev.164.1776,
	title = {Event Horizons in Static Vacuum Space-Times},
	author = {Israel, Werner},
	journal = {Phys. Rev.},
	volume = {164},
	issue = {5},
	pages = {1776--1779},
	numpages = {0},
	year = {1967},
	month = {Dec},
	publisher = {American Physical Society},
	doi = {10.1103/PhysRev.164.1776},
	url = {https://link.aps.org/doi/10.1103/PhysRev.164.1776}
}

@article{israel1968event,
	title={Event horizons in static electrovac space-times},
	author={Israel, Werner},
	journal={Communications in Mathematical Physics},
	volume={8},
	number={3},
	pages={245--260},
	year={1968},
	publisher={Springer}
}

@article{mazur1982proof,
	title={Proof of uniqueness of the Kerr-Newman black hole solution},
	author={Mazur, Pawel O},
	journal={Journal of Physics A: Mathematical and General},
	volume={15},
	number={10},
	pages={3173},
	year={1982},
	publisher={IOP Publishing}
}

@article{mazur1984black,
	title={Black hole uniqueness from a hidden symmetry of Einstein's gravity},
	author={Mazur, Pawe{\l} Oskar},
	journal={General relativity and gravitation},
	volume={16},
	number={3},
	pages={211--215},
	year={1984},
	publisher={Springer}
}

@article{PhysRevLett.26.331,
	title = {Axisymmetric Black Hole Has Only Two Degrees of Freedom},
	author = {Carter, B.},
	journal = {Phys. Rev. Lett.},
	volume = {26},
	issue = {6},
	pages = {331--333},
	numpages = {0},
	year = {1971},
	month = {Feb},
	publisher = {American Physical Society},
	doi = {10.1103/PhysRevLett.26.331},
	url = {https://link.aps.org/doi/10.1103/PhysRevLett.26.331}
}

@article{hawking1972black,
	title={Black holes in general relativity},
	author={Hawking, Stephen W},
	journal={Communications in Mathematical Physics},
	volume={25},
	number={2},
	pages={152--166},
	year={1972},
	publisher={Springer}
}

@incollection{hawking1973event,
author= {Hawking, Stephen W.},
title= {The Event Horizon},
booktitle = {Black Holes (Les Astres Occlus)},
editor= {DeWitt-Morette, Cécile and DeWitt, Bryce S.},
publisher = {Gordon and Breach},
address = {New York},
year= {1973},
pages= {1--56},
note= {Proceedings of the 23rd Session of the École d'Été de Physique Théorique, Les Houches, France, 1972}
}

@article{bardeen1973four,
	title={The four laws of black hole mechanics},
	author={Bardeen, James M and Carter, Brandon and Hawking, Stephen W},
	journal={Communications in mathematical physics},
	volume={31},
	number={2},
	pages={161--170},
	year={1973},
	publisher={Springer}
}

@article{PhysRevLett.34.905,
	title = {Uniqueness of the Kerr Black Hole},
	author = {Robinson, D. C.},
	journal = {Phys. Rev. Lett.},
	volume = {34},
	issue = {14},
	pages = {905--906},
	numpages = {0},
	year = {1975},
	month = {Apr},
	publisher = {American Physical Society},
	doi = {10.1103/PhysRevLett.34.905},
	url = {https://link.aps.org/doi/10.1103/PhysRevLett.34.905}
}

@book{hawking2023large,
	title={The large scale structure of space-time},
	author={Hawking, Stephen W and Ellis, George FR},
	year={2023},
	publisher={Cambridge university press}
}

@article{PhysRevLett.88.101101,
	title = {A Rotating Black Ring Solution in Five Dimensions},
	author = {Emparan, Roberto and Reall, Harvey S.},
	journal = {Phys. Rev. Lett.},
	volume = {88},
	issue = {10},
	pages = {101101},
	numpages = {4},
	year = {2002},
	month = {Feb},
	publisher = {American Physical Society},
	doi = {10.1103/PhysRevLett.88.101101},
	url = {https://link.aps.org/doi/10.1103/PhysRevLett.88.101101}
}

@article{Barnich:2015dvt,
    author = "Barnich, Glenn and Troessaert, C{\'e}dric and Tempo, David and Troncoso, Ricardo",
    title = "{Asymptotically locally flat spacetimes and dynamical nonspherically-symmetric black holes in three dimensions}",
    eprint = "1512.05410",
    archivePrefix = "arXiv",
    primaryClass = "hep-th",
    reportNumber = "CECS-PHY-15-10",
    doi = "10.1103/PhysRevD.93.084001",
    journal = "Phys. Rev. D",
    volume = "93",
    number = "8",
    pages = "084001",
    year = "2016"
}

@article{Alkac:2016xlr,
    author = {Alka{\c{c}}, G{\"o}khan and Kilicarslan, Ercan and Tekin, Bayram},
    title = "{Asymptotically flat black holes in 2+1 dimensions}",
    eprint = "1601.06696",
    archivePrefix = "arXiv",
    primaryClass = "hep-th",
    doi = "10.1103/PhysRevD.93.084003",
    journal = "Phys. Rev. D",
    volume = "93",
    number = "8",
    pages = "084003",
    year = "2016"
}

@article{Dias:2011at,
    author = "Dias, Oscar J. C. and Horowitz, Gary T. and Santos, Jorge E.",
    title = "{Black holes with only one Killing field}",
    eprint = "1105.4167",
    archivePrefix = "arXiv",
    primaryClass = "hep-th",
    doi = "10.1007/JHEP07(2011)115",
    journal = "JHEP",
    volume = "07",
    pages = "115",
    year = "2011"
}

@article{Sheikh-Jabbari:2016npa,
    author = "Sheikh-Jabbari, M. M. and Yavartanoo, H.",
    title = "{Horizon Fluffs: Near Horizon Soft Hairs as Microstates of Generic AdS3 Black Holes}",
    eprint = "1608.01293",
    archivePrefix = "arXiv",
    primaryClass = "hep-th",
    reportNumber = "IPM-P-2016-",
    doi = "10.1103/PhysRevD.95.044007",
    journal = "Phys. Rev. D",
    volume = "95",
    number = "4",
    pages = "044007",
    year = "2017"
}

@article{Stotyn:2011ns,
    author = "Stotyn, Sean and Park, Miok and McGrath, Paul and Mann, Robert B.",
    title = "{Black Holes and Boson Stars with One Killing Field in Arbitrary Odd Dimensions}",
    eprint = "1110.2223",
    archivePrefix = "arXiv",
    primaryClass = "hep-th",
    doi = "10.1103/PhysRevD.85.044036",
    journal = "Phys. Rev. D",
    volume = "85",
    pages = "044036",
    year = "2012"
}

@book{Hartnoll:2018xxg,
    author = "Hartnoll, Sean A. and Lucas, Andrew and Sachdev, Subir",
    title = "{Holographic Quantum Matter}",
    isbn = "978-0-262-03843-0",
    publisher = "MIT Press",
    year = "2018"
}

@article{Hawking:2016msc,
    author = "Hawking, Stephen W. and Perry, Malcolm J. and Strominger, Andrew",
    title = "{Soft Hair on Black Holes}",
    eprint = "1601.00921",
    archivePrefix = "arXiv",
    primaryClass = "hep-th",
    doi = "10.1103/PhysRevLett.116.231301",
    journal = "Phys. Rev. Lett.",
    volume = "116",
    number = "23",
    pages = "231301",
    year = "2016"
}

@article{Donnay:2023mrd,
    author = "Donnay, Laura",
    title = "{Celestial holography: An asymptotic symmetry perspective}",
    eprint = "2310.12922",
    archivePrefix = "arXiv",
    primaryClass = "hep-th",
    doi = "10.1016/j.physrep.2024.04.003",
    journal = "Phys. Rept.",
    volume = "1073",
    pages = "1--41",
    year = "2024"
}

@article{Erices:2019onl,
    author = "Erices, Cristi{\'a}n and Riquelme, Miguel and Rodr{\'\i}guez, Pablo",
    title = "{BTZ black hole with Korteweg{\textendash}de Vries-type boundary conditions: Thermodynamics revisited}",
    eprint = "1907.13026",
    archivePrefix = "arXiv",
    primaryClass = "hep-th",
    doi = "10.1103/PhysRevD.100.126026",
    journal = "Phys. Rev. D",
    volume = "100",
    number = "12",
    pages = "126026",
    year = "2019"
}

@article{Adami:2025pfk,
    author = "Adami, Hamed and Latifi, Anouchah",
    title = "{Integrability in three-dimensional gravity: eigenfunction-forced KdV flows}",
    eprint = "2510.10519",
    archivePrefix = "arXiv",
    primaryClass = "hep-th",
    doi = "10.1007/JHEP07(2026)120",
    journal = "JHEP",
    volume = "07",
    pages = "120",
    year = "2026"
}

@article{Banados:1992wn,
    author = "Banados, Maximo and Teitelboim, Claudio and Zanelli, Jorge",
    title = "{The Black hole in three-dimensional space-time}",
    eprint = "hep-th/9204099",
    archivePrefix = "arXiv",
    reportNumber = "PRINT-92-0151 (CHILE), IASSNS-HEP-92-29",
    doi = "10.1103/PhysRevLett.69.1849",
    journal = "Phys. Rev. Lett.",
    volume = "69",
    pages = "1849--1851",
    year = "1992"
}

@article{Banados:1992gq,
    author = "Banados, Maximo and Henneaux, Marc and Teitelboim, Claudio and Zanelli, Jorge",
    title = "{Geometry of the (2+1) black hole}",
    eprint = "gr-qc/9302012",
    archivePrefix = "arXiv",
    reportNumber = "IASSNS-HEP-92-81",
    doi = "10.1103/PhysRevD.48.1506",
    journal = "Phys. Rev. D",
    volume = "48",
    pages = "1506--1525",
    year = "1993",
    note = "[Erratum: Phys.Rev.D 88, 069902 (2013)]"
}

@article{Banados:2004nr,
    author = "Banados, Maximo and Caro, Rodrigo",
    title = "{Holographic ward identities: Examples from 2+1 gravity}",
    eprint = "hep-th/0411060",
    archivePrefix = "arXiv",
    doi = "10.1088/1126-6708/2004/12/036",
    journal = "JHEP",
    volume = "12",
    pages = "036",
    year = "2004"
}

@incollection{fefferman1985elie,
author= {Fefferman, Charles and Graham, C. Robin},
title = {Conformal Invariants},
booktitle = {\'Elie Cartan et les math\'ematiques d'aujourd'hui -- Lyon, 25--29 juin 1984},
series = {Ast\'erisque},
number= {S131},
pages = {95--116},
publisher = {Soci\'et\'e Math\'ematique de France},
year= {1985}
}

@book{Fefferman:2007rka,
    author    = {Fefferman, Charles and Graham, C. Robin},
    title     = {The Ambient Metric},
    series    = {Annals of Mathematics Studies},
    volume    = {178},
    publisher = {Princeton University Press},
    address   = {Princeton, NJ},
    year      = {2011},
    doi       = {10.1515/9781400840588},
    isbn      = {9780691153131}
}

@article{Ciambelli:2019bzz,
    author = "Ciambelli, Luca and Leigh, Robert G.",
    title = "{Weyl Connections and their Role in Holography}",
    eprint = "1905.04339",
    archivePrefix = "arXiv",
    primaryClass = "hep-th",
    reportNumber = "CPHT-RR020.052019",
    doi = "10.1103/PhysRevD.101.086020",
    journal = "Phys. Rev. D",
    volume = "101",
    number = "8",
    pages = "086020",
    year = "2020"
}

@article{Jia:2021hgy,
    author = "Jia, Weizhen and Karydas, Manthos",
    title = "{Obstruction tensors in Weyl geometry and holographic Weyl anomaly}",
    eprint = "2109.14014",
    archivePrefix = "arXiv",
    primaryClass = "hep-th",
    doi = "10.1103/PhysRevD.104.126031",
    journal = "Phys. Rev. D",
    volume = "104",
    number = "12",
    pages = "126031",
    year = "2021"
}

@article{Ciambelli:2023ott,
    author = "Ciambelli, Luca and Delfante, Arnaud and Ruzziconi, Romain and Zwikel, C{\'e}line",
    title = "{Symmetries and charges in Weyl-Fefferman-Graham gauge}",
    eprint = "2308.15480",
    archivePrefix = "arXiv",
    primaryClass = "hep-th",
    doi = "10.1103/PhysRevD.108.126003",
    journal = "Phys. Rev. D",
    volume = "108",
    number = "12",
    pages = "126003",
    year = "2023"
}

@article{Arenas-Henriquez:2024ypo,
    author = "Arenas-Henriquez, Gabriel and Diaz, Felipe and Rivera-Betancour, David",
    title = "{Generalized Fefferman-Graham gauge and boundary Weyl structures}",
    eprint = "2411.12513",
    archivePrefix = "arXiv",
    primaryClass = "hep-th",
    doi = "10.1007/JHEP02(2025)007",
    journal = "JHEP",
    volume = "02",
    pages = "007",
    year = "2025"
}

@article{Skenderis:1999nb,
    author = "Skenderis, Kostas and Solodukhin, Sergey N.",
    title = "{Quantum effective action from the AdS / CFT correspondence}",
    eprint = "hep-th/9910023",
    archivePrefix = "arXiv",
    reportNumber = "SPIN-1999-23",
    doi = "10.1016/S0370-2693(99)01467-7",
    journal = "Phys. Lett. B",
    volume = "472",
    pages = "316--322",
    year = "2000"
}

@article{Banados:1994tn,
    author = "Banados, Maximo",
    title = "{Global charges in Chern-Simons field theory and the (2+1) black hole}",
    eprint = "hep-th/9405171",
    archivePrefix = "arXiv",
    reportNumber = "IMPERIAL-TP-93-94-40",
    doi = "10.1103/PhysRevD.52.5816",
    journal = "Phys. Rev. D",
    volume = "52",
    pages = "5816--5825",
    year = "1996"
}

@article{Banados:1998ta,
    author = "Banados, Maximo and Brotz, Thorsten and Ortiz, Miguel E.",
    title = "{Boundary dynamics and the statistical mechanics of the (2+1)-dimensional black hole}",
    eprint = "hep-th/9802076",
    archivePrefix = "arXiv",
    reportNumber = "IMPERIAL-TP-97-98-23, DFTUZ-98-05",
    doi = "10.1016/S0550-3213(99)00069-3",
    journal = "Nucl. Phys. B",
    volume = "545",
    pages = "340--370",
    year = "1999"
}

@inproceedings{Troessaert:2015syk,
    author = "Troessaert, C{\'e}dric and Tempo, David and Troncoso, Ricardo",
    title = "{Asymptotically flat black holes and gravitational waves in three-dimensional massive gravity}",
    booktitle = "{8th Aegean Summer School}: {Gravitational Waves: From Theory to Observations}",
    eprint = "1512.09046",
    archivePrefix = "arXiv",
    primaryClass = "hep-th",
    reportNumber = "CECS-PHY-15-11",
    month = "12",
    year = "2015"
}

@article{Sheikh-Jabbari:2014nya,
    author = "Sheikh-Jabbari, M. M. and Yavartanoo, Hossein",
    title = "{On quantization of AdS$_{3}$ gravity I: semi-classical analysis}",
    eprint = "1404.4472",
    archivePrefix = "arXiv",
    primaryClass = "hep-th",
    doi = "10.1007/JHEP07(2014)104",
    journal = "JHEP",
    volume = "07",
    pages = "104",
    year = "2014"
}

@article{Balog:1997zz,
    author = "Balog, J. and Feher, L. and Palla, L.",
    title = "{Coadjoint orbits of the Virasoro algebra and the global Liouville equation}",
    eprint = "hep-th/9703045",
    archivePrefix = "arXiv",
    reportNumber = "BONN-TH-97-02, ITP-BUDAPEST-526",
    doi = "10.1142/S0217751X98000147",
    journal = "Int. J. Mod. Phys. A",
    volume = "13",
    pages = "315--362",
    year = "1998"
}

\end{document}